# Discovering hierarchical governing equations for community-structured complex network dynamics with LLM agents

Jinlei Zhang[1,6,*], Zixuan Zhang[1], Hao Qiu[1], Feng Chen[2], Jiateng Yin[1], Xiaoyong Yan[1], Xiaolei Ma[3], Hao Sun[4], Jianxi Gao[5], Lixing Yang[1,**] and Ziyou Gao[1,***]

[1]School of Systems Science, Beijing Jiaotong University, Beijing, China

[2]School of Civil Engineering, Beijing Jiaotong University, Beijing, China.

[3]School of Transportation Science and Engineering, Beihang University, Beijing, China

[4]Gaoling School of Artificial Intelligence, Renmin University of China, Beijing, China

[5]Department of Computer Science, Rensselaer Polytechnic Institute, Troy, NY, US.

[6]Lead contact

*Correspondence: zhangjinlei@bjtu.edu.cn

**Correspondence: lxyang@bjtu.edu.cn

***Correspondence: zygao@bjtu.edu.cn

**SUMMARY**

Governing equations provide an interpretable description of the mechanisms underlying complex network dynamics, yet their discovery from observational data remains challenging when shared community-level dynamics coexist with individual-level variations and the relevant functional structures may be absent from the initial candidate library or not explicitly specified in advance. Here we formulate this task as a hierarchical law-discovery problem and introduce SAGE, a large language model-guided self-reflective agent framework for discovering community-shared governing laws and individual-specific dynamical variations. SAGE organizes equation discovery into an iterative Library–Discovery–Review process that first identifies shared community-level structures and then resolves node-specific residual dynamics, while adaptively refining the candidate space to introduce missing dynamical terms when needed. This hierarchical and interpretable formulation enables SAGE to distinguish shared community-level dynamics from individual-level variations and to reveal the mechanisms underlying complex network dynamics. Across two synthetic networks and four real-world networks spanning mobile communication, brain electrophysiology, power grids and urban rail transit, SAGE recovers compact and interpretable governing equations, reproduces observed system dynamics and outperforms representative equation-discovery baselines. The framework remains robust under incomplete observations and can transfer community-level governing laws to nodes with entirely missing observations. These results establish a hierarchical and adaptive approach for disentangling shared and individual dynamical mechanisms in community-structured complex networks.

**INTRODUCTION**

Inferring the governing laws of complex networks, such as ecological [1,2], neuronal [3,4], social [5,6], and transportation [7,8] networks, is fundamental to understanding these systems. Many complex networks exhibit clear community structures and individual-level variations in their dynamics. These networks are typically characterized by the coexistence of community-level dynamics and individual-level dynamics. Community-level dynamics [9–11] describe dynamical behaviors shared among nodes within the same community, giving rise to similar community-level dynamical patterns. Individual-level dynamics [12–14], on the other hand, reflect dynamical variations among individual nodes within the same community. Using a single universal governing equation to describe the dynamics of all nodes in a community-structured network may overlook dynamical differences across communities. In contrast, assigning a separate governing equation to every node ignores shared dynamical structure, typically requires more data, and reduces interpretability. The central challenge is therefore to determine which dynamical mechanisms are shared among nodes within the same community and which represent node-specific deviations. This motivates a hierarchical representation in which the governing dynamics of each node are decomposed into community-level dynamical components shared within the community and individual-level components that capture node-specific deviations. This decomposition is relevant to many complex real-world networks. For example, in neuronal networks, neurons of the same type or within the same brain region are often assumed to share similar core intrinsic dynamics. However, differences in ion-channel composition or expression levels can introduce additional dynamics through parameter changes or additional current terms. These differences can lead to node-specific variations in neuronal activity. Modeling both community-level and individual-level dynamics thus enables a more complete description of community-level dynamical patterns and node-specific variations in complex networks.

Conventional approaches derive governing equations for complex networks from first principles or domain knowledge–based derivations [15]. However, these methods often require strong prior assumptions and predefined model structures [16–18]. Such requirements limit their use in real-world networks. These networks are often highly nonlinear, and their underlying mechanisms may remain poorly understood. The growing availability of large-scale spatiotemporal data has enabled data-driven discovery of governing laws in complex networks. For example, symbolic regression methods [19–23] can infer governing equations of complex systems using combinations of predefined variables and operators to fit observational data. Beyond networked systems, recent graph-based equation discovery has further demonstrated the potential to identify compact and interpretable physical laws directly from experimental data [24]. However, equation-discovery methods that search over combinations of variables and operators can face a rapidly growing search space, leading to high computational costs and potential overfitting. SINDy [25] builds a library of candidate functions and applies sparse regression to identify dominant terms. As an extension of this sparse-library paradigm, PDE-FIND [26] extends this sparse-library paradigm to the discovery of partial differential equations. These sparse-library methods formulate equation discovery as a regularized regression problem that balances model complexity and fitting accuracy. SINDy and its variants have been applied to a wide range of domains, including physical [27,28], biological [29], and engineering systems [30,31]. More recent studies have combined equation discovery with deep learning, reinforcement learning, and neural symbolic search [32–38]. Despite these advances, sparse-library approaches generally rely on a predefined candidate function space. When relevant dynamical terms are absent from the initial library, existing methods often lack a systematic way to revise the candidate space based on model failures. As a result, missing candidate functions may remain unaddressed, leading to incomplete or inaccurate governing equations. This is especially problematic in community-structured networks, where residual dynamics may reflect either true node-specific behavior or missing terms in the candidate library.

For complex networks, an additional challenge arises because node dynamics reflect both intrinsic mechanisms and network interactions [39]. The Two-Phase framework enables autonomous inference of deterministic complex network dynamics from incomplete and noisy observations [40]. LaGNA further extends interpretable equation discovery to stochastic complex systems by modeling intrinsic dynamics, network interactions, and stochastic diffusion, thereby learning interpretable stochastic differential equations from observational data [41]. Despite this progress, many existing methods use a shared functional form across all nodes. Such a global model may overlook dynamical differences between communities and node-specific deviations within each community. Conversely, fitting an independent equation for every node ignores shared dynamical structure and typically requires more data. Therefore, existing methods do not directly address the joint identification of community-level and individual-level dynamics in community-structured complex networks.

Recent studies have used large language models (LLMs) to guide symbolic and scientific search. LaSR uses an LLM to discover and evolve reusable concepts for symbolic regression [42]. D3 employs multiple agents to propose, evaluate, and iteratively refine interpretable dynamical models [43]. Reflection-based methods also show that language agents can improve their output through feedback without updating model parameters. Reflexion stores linguistic reflections in episodic memory [44], while Self-Refine improves output through repeated feedback and revision [45]. These studies show that LLMs can support iterative model refinement through feedback. However, governing-equation discovery in community-structured complex networks still faces two coupled challenges: distinguishing shared community-level dynamics from node-specific variations, and revising the candidate function space when important dynamical terms are missing. A unified framework is therefore needed to address both challenges within the same discovery process.

To address these challenges, we formulate governing-equation discovery in community-structured complex networks as a hierarchical law-discovery problem, in which node dynamics are decomposed into community-level governing laws shared among related nodes and individual-level components that capture node-specific deviations. We develop a Large Language Model–guided Self-reflective Agent framework for Governing Equation discovery (SAGE) to solve this problem when the candidate function space may initially be incomplete. SAGE organizes equation discovery into an iterative candidate-space construction, equation identification, and evaluation process. The Library Agent constructs and adaptively refines the candidate function library by integrating network observations with diagnostic feedback. The Discovery Agent first identifies community-level equation structures and then analyzes the remaining residual dynamics to determine whether additional individual-level terms are required, while Pareto selection balances equation complexity and model accuracy. The Review Agent evaluates candidate equations through derivative consistency, coefficient stability, and state prediction, providing feedback when the inferred dynamics remain incomplete or inconsistent. When structural mismatch is detected, failed review criteria and residual patterns are converted into targeted repair instructions that guide the next round of candidate-library refinement. Importantly, the LLM does not directly generate equations or estimate numerical parameters; instead, it assists in identifying missing dynamical mechanisms and refining the candidate function space, whereas candidate equation generation and coefficient estimation remain handled by numerical equation-discovery methods. In this way, SAGE combines hierarchical dynamics discovery with adaptive candidate-space refinement in an interpretable and iterative scientific discovery process:

(1) We formulate governing-equation discovery in community-structured complex networks as a hierarchical law-discovery problem and propose SAGE, a unified Large Language Model–guided Self-reflective Agent framework for Governing Equation discovery. Rather than assuming either a single shared equation or fully independent node-wise equations, the framework jointly identifies community-level dynamics shared among related nodes and individual-level dynamics that capture node-specific deviations. The iterative process is interpretable, reviewable, repairable, and traceable.

(2) We treat the candidate function library as an adaptive rather than fixed search space. When a candidate equation fails a repairable review criterion, SAGE uses failed review checks and residual patterns to identify potentially missing dynamical mechanisms and guide targeted expansion or revision of the candidate library under explicit complexity constraints. This process allows the framework to refine its search without updating the LLM parameters.

(3) We conduct extensive experiments on two synthetic and four real-world community-structured complex networks from different domains. The results demonstrate the interpretability, predictive accuracy, and robustness of SAGE. SAGE also outperforms representative equation-discovery baselines and accurately infers governing equations even under incomplete observation settings. Furthermore, the discovered community-level dynamics can be transferred to nodes with entirely missing observations, demonstrating the value of explicitly identifying shared governing laws in data-limited complex networks.

## RESULTS

### *Overview of the SAGE framework*

Complex networks often exhibit both community-level and individual-level dynamics **(Fig. 1)**. Considering a complex network with N nodes and K communities, the dynamics of each node can be described by a hierarchical governing equation composed of community-level dynamical terms together with additional individual-level dynamical terms:

$$\frac{\mathrm{d}\boldsymbol{x}_{\mathrm{i}}(t)}{\mathrm{d}t} = \boldsymbol{S}_k(\boldsymbol{x}_i(t), \boldsymbol{A}) + \boldsymbol{H}_i(\boldsymbol{x}_i(t), \boldsymbol{A}),\ i \in community\ k \quad (1)$$

where $\boldsymbol{x}_i(t) \in \mathbb{R}^d$ denotes the d-dimensional state vector of node $i$ at time $t$ and $\boldsymbol{A} \in \mathbb{R}^{N \times N}$ represents the network adjacency matrix. The node dynamics $\frac{\mathrm{d}x_{\mathrm{i}}(t)}{\mathrm{d}t} \in \mathbb{R}^d$ consist of both community-level and individual-level components. $\boldsymbol{S}_k(\boldsymbol{x}_i(t), \boldsymbol{A}) \in \mathbb{R}^d$ denotes the community-level dynamics for community $k$. All nodes in community $k$ share the same $\boldsymbol{S}_k$, which captures their shared dynamical behavior. $\boldsymbol{H}_i(\boldsymbol{x}_i(t), \boldsymbol{A}) \in \mathbb{R}^d$ denotes the individual-level dynamics of the node $i$. It captures unique dynamic behaviors caused by intrinsic attributes. Each node has its own individual-level function $\boldsymbol{H}_i$. These functions capture variation across nodes.

In real-world complex networks, prior knowledge of network dynamics ($\boldsymbol{S}_k$ and $\boldsymbol{H}_i$) is often limited or unknown. Our goal is to automatically infer the hierarchical governing equations of complex networks from observational data, including both community-level dynamics and individual-level dynamics. We assume that governing equations usually contain only a few key terms, which can be selected from a large library of candidate functions. For simplicity, we consider one dimension $\boldsymbol{x}_{\mathrm{i}}(t)$ of node $i$ in community $k$. Then, Equation 1 can be written as:

$$\frac{\mathrm{d}x_i(t)}{\mathrm{d}t} = S_k\big(x_i(t)\big) + H_i\big(x_i(t)\big) = \boldsymbol{\Theta}^{(r)}(x_i(t))\boldsymbol{\xi}_k^{S(r)} + \boldsymbol{\Theta}^{(r)}(x_i(t))\boldsymbol{\xi}_i^{H(r)} \quad (2)$$

where $\boldsymbol{\Theta}^{(r)}(x_i(t))$ denotes the adaptive candidate-function library at search round $r$, including polynomial, trigonometric, exponential, and interaction terms (see Table S2). The library is initialized with these common function families and can be adaptively expanded by the Library Agent during equation discovery under explicit complexity constraints. $M_r$ denotes the number of candidate functions in the current library at search round $r$. $\boldsymbol{\xi}_k^{S(r)} = \left[\xi_{k,1}^{S(r)}, \cdots, \xi_{k,M}^{S(r)}\right]^T \in \mathbb{R}^{M_r}$ denotes the coefficient vector of the community-level dynamics shared by nodes in community $k$. $\boldsymbol{\xi}_i^{H(r)} = \left[\xi_{i,1}^{H(r)}, \cdots, \xi_{i,M}^{H(r)}\right]^T \in \mathbb{R}^{M_r}$ denotes the coefficient vector of the individual-level dynamics of node $i$. The superscript denotes the search round in which the candidate library and corresponding coefficients are obtained. Accordingly, hierarchical governing-equation discovery can be formulated as a joint sparse estimation problem for the community-level coefficients $\boldsymbol{\xi}_k^{S(r)}$ and the individual-level coefficients $\boldsymbol{\xi}_i^{H(r)}$.

To solve this hierarchical estimation problem, SAGE identifies community-level and individual-level dynamics from node-state observations and network structure through a unified hierarchical procedure **(Fig. 1b)**. The framework comprises three agents: the Library Agent, the Discovery Agent, and the Review Agent.

The Library Agent preprocesses the observations, estimates temporal derivatives and splits the data into training, validation and test sets. It then builds an initial candidate function library with self-dynamics, polynomials, trigonometric functions, saturation terms, temporal terms and topology-aware interactions. Each candidate term has an explicit mathematical form linked to a possible dynamical mechanism, which supports physical interpretation of the inferred equations. The library is adaptive rather than fixed. When residuals suggest missing oscillatory, saturation or nonlinear interaction terms, etc., the Library Agent expands the relevant function families under explicit complexity constraints. For each node, the library matrix $\mathbf{\Theta}^{(r)}(x_i(t))$ provides a common basis for estimating community-level and individual-level terms.

The Discovery Agent searches for equations that explain the observed node dynamics. It is designed to be independent of any specific numerical solver and uses interchangeable numerical equation-discovery methods to generate candidate equation structures and estimate their coefficients. In this study, we instantiate the Discovery Agent with three representative sparse regression methods: bootstrap sequentially thresholder least squares, cross-validated Lasso, and ridge regression under multiple threshold settings. These numerical methods are not fixed components of SAGE and can be replaced or extended by other compatible equation discovery methods without changing the overall framework. Rather than relying on a single solution, the agent compares candidate equations based on validation error and equation complexity. It then selects a Pareto-optimal model that balances accuracy and simplicity. The search proceeds at two levels. At the community level, the Discovery Agent identifies dynamical terms shared by nodes within the same community and estimates community-specific coefficients. At the individual level, it analyzes the remaining unexplained dynamics and determines whether additional node-specific dynamical terms are required.

The Review Agent serves as an independent validation and decision layer. It evaluates each selected equation on the validation set using three complementary checks: derivative fit, coefficient stability, and state prediction. The agent also examines whether the selected terms have clear physical meaning. Based on these results, the Review Agent returns one of three decisions: ACCEPT, REPAIR or REJECT. ACCEPT indicates that the equation meets all review criteria. The agent then reports its structure, coefficients, predictive performance and physical interpretation. REPAIR indicates that the equation is incomplete or unstable. The agent identifies the failed checks, converts them into a structured repair plan and sends targeted instructions to the Library Agent. These instructions specify which function families should be revised or expanded in the next search round. REJECT ends the search when the available data cannot support a reliable equation or when the maximum number of iterations is reached and then provides a report that explains the remaining problems.

SAGE repeats the Library–Discovery–Review cycle until an equation passes the review criteria or a stopping criterion is reached. Throughout this process, the framework records the candidate libraries, inferred equations, review results, and repair actions. These records make the discovery process transparent and traceable. Once accepted, the selected community-level and individual-level terms, together with their coefficients, define the governing equation for each node. The inferred node dynamics are expressed as follows **(Fig. 1c)**:

$$\frac{\mathrm{d}\hat{x}_i(t)}{\mathrm{d}t} = \mathbf{\Theta}^{(r)}(x_i(t))\hat{\boldsymbol{\xi}}_k^{S(r)} + \mathbf{\Theta}^{(r)}(x_i(t))\hat{\boldsymbol{\xi}}_i^{H(r)} \tag{3}$$

To further validate whether the discovered equation can reproduce the observed system evolution, the inferred dynamics numerically evolved over a time interval $\Delta t$:

$$\hat{x}_i(t+\Delta t) = x_i(t) + \frac{d\hat{x}_i(t)}{dt}\Delta t \tag{4}$$

here, $\hat{x}_i(t)$ denotes the inferred state of node $\hat{x}_i(t)$, and $\Delta t$ represents the prediction interval. The superscript $r$ denotes the search round at which the candidate-function library and the corresponding coefficient vectors are obtained. $\hat{\boldsymbol{\xi}}_k^{S(r)}$ and $\hat{\boldsymbol{\xi}}_i^{H(r)}$ denote the corresponding community-level and individual-level coefficient vectors, respectively. Equation 3 represents the discovered governing dynamics, whereas Equation 4 is used as a validation step to assess whether these dynamics can reproduce the observed node-state trajectories.

Further details are provided in Methods and Document S1, section 1.

SAGE is a multi-agent framework for governing equation discovery that organizes the inference process into an iterative Library–Discovery–Review loop. The Library Agent prepares the numerical inputs and constructs and updates the candidate function library, while also identifying community structures from node observations when prior community information is unavailable.

The Discovery Agent generates and selects sparse equation structures, and the Review Agent evaluates the selected equations and guides subsequent repair when necessary. The main notation and variable dimensions are summarized in Table S1.

In this study, we consider a complex network with multiple communities, as shown in **Figure 2a**. The framework takes node state observations and the adjacency matrix as primary inputs. Community assignments are used for subsequent hierarchical dynamics discovery, where available community labels are directly adopted; otherwise, they are inferred by the Library Agent from node state observations.

### *Inferring synthetic complex network dynamics*

To evaluate the effectiveness and robustness of SAGE, we construct two synthetic complex networks: a single-dynamics network and a multi-dynamics network. Both networks simultaneously exhibit community-level dynamics and individual-level dynamics. Specifically, we first generate network structures using the Stochastic Block Model (SBM) [46] and then simulate network dynamics on the generated topology to obtain node-state observations. In our experiments, the single-dynamics network highlights inter-community differences by assigning distinct community-level coefficients to different communities while sharing the same underlying dynamics. In contrast, the multi-dynamics network captures inter-community differences by applying different types of network dynamics to different communities.

In the single-dynamics network, all nodes follow the Lotka–Volterra dynamics [47,48]:

$$\begin{cases} \frac{\mathrm{d}x_i}{\mathrm{d}t} = a_k x_i - b_k x_i y_i + c_k \sum_{j \in N_i} x_j + \phi_i \sin(1.5t) \cdot x_i \\ \frac{\mathrm{d}y_i}{\mathrm{d}t} = d_k x_i y_i - e_k y_i + \varphi_i \sin(1.5t) \cdot y_i \end{cases} \tag{5}$$

where $a_k, b_k, c_\mathrm{k}, d_k, e_k$ are community-level coefficients for community $k$. Nodes within the same community $k$ share these coefficients, which characterize their shared dynamical behavior. In contrast, different communities are assigned to distinct coefficients, highlighting dynamical differences across communities. $\phi_i, \varphi_i$ are individual-level coefficients for node $i$ and represent individual-level variations. Each node has distinct coefficients that capture its individual dynamical behavior.

In the multi-dynamics network, nodes follow three dynamical systems: the Lorenz [49], Rössler [50], and Chen [51] systems. This network highlights stronger dynamics differences across communities:

$$Lorenz \begin{cases} \frac{\mathrm{d}x_i}{\mathrm{d}t} = a_1(y_i - x_i) + \phi_i x_i \\ \frac{\mathrm{d}y_i}{\mathrm{d}t} = x_i(b_1 - z_i) - y_i + \varphi_i y_i \\ \frac{\mathrm{d}z_i}{\mathrm{d}t} = x_i y_i - c_1 z_i + \psi_i z_i \end{cases} \quad Chen \begin{cases} \frac{\mathrm{d}x_i}{\mathrm{d}t} = a_3(y_i - x_i) + \phi_i x_i \\ \frac{\mathrm{d}y_i}{\mathrm{d}t} = -x_i z_i + c_3 y_i + \varphi_i y_i \\ \frac{\mathrm{d}z_i}{\mathrm{d}t} = x_i y_i - b_3 z_i + \psi_i z_i \end{cases}$$
$$Rossler \begin{cases} \frac{\mathrm{d}x_i}{\mathrm{d}t} = -y_i - z_i + \phi_i z_i \\ \frac{\mathrm{d}y_i}{\mathrm{d}t} = x_i + a_2 y_i + \varphi_i y_i \\ \frac{\mathrm{d}z_i}{\mathrm{d}t} = b_2 + z_i(x_i - c_2) + \psi_i \end{cases} \tag{6}$$

where $a, b, c$ are community-level coefficients. Each node follows a governing equation derived from one of the three systems-Lorenz, Rössler, or Chen. $\phi_i, \varphi_i, \psi_i$ are individual-level coefficients for node $i$.

**Figures 3 and 4** demonstrate that SAGE can accurately infer governing equations for complex networks with coupled community-level and individual-level dynamics. For the single-dynamics network **(Fig. 3)**, we first examine whether SAGE can identify the relevant dynamical terms from the adaptive function library. The Discovery Agent searches the candidate function library and identifies the necessary terms to describe the system dynamics **(Fig. 3a)**. After review and refinement, the selected terms correspond to the main components of the governing equations. The inferred equations reproduce the true phase portraits, demonstrating that SAGE captures the underlying governing dynamics of the system **(Fig. 3b)**. The inferred derivatives closely match the true derivatives **(Fig. 3c)**, while the inferred equations accurately reproduce node states **(Fig. 3d)**. The quantitative comparison across overall, community-level, and individual-level dynamics further confirms the accuracy of the inferred equations, with $R^2$ values reaching 0.99 **(Fig. 3e)**. Finally, the inferred equations recover both shared community-level terms and individual-level terms, with relative coefficient errors below 2.71% **(Fig. 3f)**.

Beyond the single-dynamics case, we further evaluate whether SAGE can distinguish different governing systems across communities. In this setting, nodes from different communities follow distinct nonlinear systems (Lorenz, Rössler, and Chen), while nodes within each community retain individual-level dynamics. The Discovery Agent identifies the functional terms required by each governing system **(Fig. 4a)**. The inferred equations successfully reproduce the phase portraits of the true systems, demonstrating that SAGE can recover distinct dynamical behaviors across different communities **(Fig. 4b)**. The inferred derivatives and node states remain consistent with the true systems **(Fig. 4c, d)**. The quantitative evaluation further confirms accurate recovery across overall, community-level, and individual-level dynamics**(Fig. 4e)**. Finally, SAGE recovers distinct governing equations across communities while preserving individual-level variations within each community**(Fig. 4f)**.

In summary, SAGE demonstrates the ability to infer governing equations and reconstruct network dynamics for both a single-dynamics network **(Fig. 3)** and a multi-dynamics network **(Fig. 4)**. In the single-dynamics network, the model accurately identifies community-level and individual-level dynamical terms, while recovering the true governing equations and dynamical trajectories. In the multi-dynamics network, although different communities follow distinct governing systems, SAGE effectively identifies the corresponding governing equations together with community-level and individual-level dynamics. These two experiments demonstrate the equation recovery capability, and equation recovery capability of SAGE in modeling networks with coupled community-level and individual-level dynamics. Detailed system descriptions and inference results for synthetic complex networks can be found in Document S1, section 2.

### *Validation of discovered governing equations in real-world networks*

To evaluate whether SAGE can uncover meaningful dynamical mechanisms in real-world networks, we apply it to four community-structured complex networks, including an urban mobile communication network [52], a brain EEG network [53], a power grid network [54], and an urban rail transit network [55]. These systems represent distinct classes of complex dynamics, including human activity-driven interactions, biological oscillations, infrastructure synchronization, and mobility propagation.

We first examine whether the discovered governing equations reveal network-specific dynamical mechanisms consistent with the intrinsic characteristics of each network. For each accepted equation, we analyze the retained dynamical terms from three perspectives: (i) the dominant mechanisms encoded by the equation structure, (ii) the correspondence between discovered terms and known network dynamics, and (iii) the separation between community-level shared mechanisms and individual-level variations. State reconstruction and short-term numerical evolution are used only as quantitative validation tools to assess whether the discovered equations can reproduce the observed network dynamics.

Detailed dataset descriptions, discovered equations, and additional validation results are provided in Document S1, section 3.

### *Urban mobile communication network*

Urban mobile communication networks exhibit recurrent temporal rhythms, nonlinear interactions, and substantial spatial heterogeneity driven by human mobility and regional activity patterns. To examine whether SAGE can recover these characteristic dynamical structures, we select Node 1 in Community 1 as a representative case. For each node, the three state variables $x_i(t)$, $y_i(t)$, and $z_i(t)$ represent Internet activity, phone-call activity, and SMS activity, respectively. The discovered equation structure and its hierarchical interpretation are summarized in **Fig. 5**, while additional interpretation results are provided in Document S1, section 3.

The discovered equations reveal three main dynamical mechanisms: state–time modulation, time-modulated cross-state coupling, and nonlinear state dependence. State–time modulation terms, such as $x\cos(w_0 t)$, $y\cos(w_0 t)$, and $z\sin(w_0 t)$, indicate that the effects of communication states vary with recurrent temporal phases. Time-modulated cross-state terms show that the evolution of one communication state can be influenced by other states under the same temporal modulation. In addition, the nonlinear term $z^2$ captures state-dependent dynamics that cannot be represented by linear responses alone. Together, these terms show that urban communication dynamics are shaped by recurrent temporal modulation, cross-state coupling, and nonlinear state dependence.

At the community level, nodes within the same community share similar state–time modulation and cross-state coupling structures, indicating common temporal and interaction mechanisms. At the individual level, node-specific terms and coefficient variations capture local departures from these shared structures, including differences in temporal responses and nonlinear state dependence. Thus, SAGE separates community-shared dynamical mechanisms from node-specific variations, providing an explicit and interpretable description of heterogeneous urban communication dynamics

### *Brain EEG network*

Brain EEG networks are representative biological complex networks composed of interacting neural channels, characterized by nonlinear oscillatory activity, functional connectivity, and dynamical variation across channels. EEG channels within functionally related communities may share common interaction patterns while retaining channel-specific deviations. Here, we analyze the governing equations discovered by SAGE to reveal the dynamical structures underlying community-level oscillatory behavior and individual variations in EEG networks. For each channel $i$, the two state variables $x_i(t)$ and $y_i(t)$ denote the real component of the filtered EEG analytic signal and its corresponding Hilbert quadrature component, respectively.

SAGE identifies the following compact governing equations for the EEG dynamics:

$$\frac{dx}{dt} = (-10.7684y + 5.7214) + (-1.3214y - 0.1758) = -12.0898y + 5.5456$$
$$\frac{dy}{dt} = (17.6635x - 6.6710) + (-0.5147x - 0.8615) = 17.1488x - 7.5325 \tag{7}$$

The discovered equations reveal two main dynamical mechanisms: cross-state coupling and constant bias. The cross-state coupling terms, $y$ and $x$, indicate that the evolution of each EEG state is governed by the other state rather than by an isolated self-dynamic term, revealing a coupled dynamical structure underlying EEG activity. Moreover, the opposite signs of these coupling terms suggest an antisymmetric coupling pattern consistent with coupled oscillatory dynamics between the real and quadrature components, rather than independent evolution of the two states. The constant terms, 5.5456 and -7.5325, introduce state-specific offsets that shift the dynamical baseline away from the origin. Together, these terms indicate that the discovered EEG dynamics are primarily characterized by linear cross-state coupling between the two analytic-signal components together with non-zero baseline offsets.

At the community level, EEG channels within the same dynamical community share similar cross-state coupling structures, indicating coordinated oscillatory behavior governed by related dynamical mechanisms. At the individual level, channel-specific coefficient variations and baseline shifts capture local deviations from these shared community dynamics. Thus, SAGE separates community-shared coupling structures from channel-specific dynamical variations, providing an explicit and interpretable description of EEG network dynamics. Detailed dataset descriptions and additional interpretation results are provided in Document S1, section 3.

***Power grid network***

Power grid networks are representative critical infrastructure systems composed of interconnected buses whose dynamics are strongly shaped by transmission-line topology and phase interactions. Variations in operating conditions and network connectivity can lead to heterogeneous dynamical behaviors across individual buses, while buses within the same dynamical community may share similar interaction structures. Here, the state variable $x_i$ represents the phase state of bus $i$.

SAGE identifies the following topology-mediated governing equation:

$$\frac{dx}{dt} = -K \sum_{j} A_{ij} \sin(x_i - x_j) \tag{8}$$

where $A_{ij}$ represents the physical transmission-line connection between buses $i$ and $j$, and $K$ denotes the fitted coupling strength. The discovered equation reveals two main dynamical mechanisms: an angle–velocity relation and topology-dependent phase coupling. The derivative $dx_i/dt$ describes the angular velocity associated with the evolution of the rotor-angle state, explicitly linking rotor-angle evolution to its rate of change. More importantly, the adjacency-weighted term $A_{ij}\sin(x_i - x_j)$ shows that the dynamics of each bus depend on phase differences with physically connected neighboring buses. The sinusoidal interaction captures the nonlinear effect of phase differences, while the adjacency matrix restricts these interactions to the physical transmission network. Together, these terms show that power-grid dynamics are governed by local rotor-angle evolution and topology-mediated phase interactions, providing a compact description of coherent electromechanical dynamics.

At the community level, buses within the same dynamical community share similar angle–velocity relations and topology-dependent phase-coupling structures, indicating common electromechanical interaction patterns shaped by network connectivity. At the individual level, node-specific coupling coefficients and local interaction strengths capture deviations from these shared community-level dynamics. Thus, SAGE separates community-shared phase-coupling mechanisms from bus-specific interaction variations, providing an explicit and interpretable description of power-grid dynamics. Detailed dataset descriptions and additional interpretation results are provided in Document S1, section 3.

### *Urban rail transit network*

Urban rail transit networks are representative transportation infrastructure systems formed by interconnected stations and passenger-flow interactions, characterized by strong temporal dependence, nonlinear demand evolution, and spatially heterogeneous travel patterns. Influenced by passenger mobility, station functionality, commuting behaviors, and network connectivity, urban rail systems exhibit both community-level flow dynamics among stations with similar urban roles and individual variations caused by local demand fluctuations and station-specific characteristics. Here, we analyze the governing equations discovered by SAGE for passenger-flow evolution and identify the dynamical mechanisms underlying temporal dependence and heterogeneous mobility patterns.

SAGE identifies the following compact governing equations for the two passenger-flow state variables:

$$\begin{aligned}\frac{\mathrm{d}x}{\mathrm{d}t} &= (-1.2599x(t-1) - 0.3318x^2 - 0.4036y(t-1)) + (-0.0130Ay) \\ \frac{\mathrm{d}y}{\mathrm{d}t} &= (-1.8289y(t-1) - 0.7135y^2 + 0.1986x(t-1)) + (1.7491x_iy^2)\end{aligned} \tag{9}$$

The discovered equations reveal four main dynamical mechanisms: short-term memory, nonlinear state dependence, network coupling, and node-specific cross-state interaction. Short-term memory is reflected by the lagged terms, such as $x(t-1)$ and $y(t-1)$, indicating that current passenger-flow changes depend on the previous time step and therefore retain temporal persistence. Nonlinear state dependence is captured by quadratic state terms, showing that the response of passenger flow to its current state is not purely linear. Network coupling is represented by terms such as $Ay$, indicating that the dynamics of a station are influenced by passenger-flow conditions in connected neighboring stations. In addition, node-specific cross-state interaction terms, such as $xy$, show that interactions between the two passenger-flow variables vary across stations and contribute to local dynamical heterogeneity. Together, these terms indicate that urban rail transit dynamics are jointly shaped by temporal memory, nonlinear state dependence, network structure, and local cross-state interactions.

At the community level, stations with similar urban functions and travel-demand patterns share common lagged, nonlinear, and network coupling terms, indicating similar passenger flow evolution mechanisms among related stations. At the individual level, node-specific cross state interaction terms together with coefficient variations capture local deviations from these shared dynamics, reflecting heterogeneous passenger flow responses across stations. Thus, SAGE separates community shared passenger flow mechanisms from station-specific dynamical variations, providing an explicit and interpretable description of urban rail transit dynamics. Detailed dataset descriptions, interpretation results, and validation evidence are provided in Document S1, section 3.

To further validate whether the discovered equations accurately describe the underlying system evolution, we evaluate their ability to reconstruct node-state trajectories on real-world networks. The states generated from the inferred equations closely match the observed trajectories across overall, community-level, and individual-level dynamics **(Fig. 6)**. The inferred equations successfully reproduce both shared dynamical patterns among nodes within the same community and individual-level variations among nodes, demonstrating that SAGE can capture hierarchical dynamical structures in real-world complex networks.

***SAGE remains robust under incomplete observations***

Real-world complex networks often suffer from incomplete observations caused by sensor failures, communication errors, or limited monitoring resources. Such missing observations make it challenging to identify the underlying governing dynamics of complex systems. To evaluate whether SAGE can recover reliable dynamical laws under incomplete observations, we conduct two experiments by introducing different missing patterns in both synthetic networks and a real-world urban rail transit network.

Similarly, we use node state prediction as an indirect evaluation criterion to examine whether the discovered governing equations can accurately characterize the underlying system dynamics. Specifically, if SAGE successfully identifies the governing equations from incomplete observations, the inferred equations should be able to reconstruct missing dynamical trajectories and accurately describe future system evolution.

In real-world networks, individual sensors may temporarily fail, or communication errors may cause missing observations at certain time steps. To simulate this scenario, we randomly remove observations from all nodes at specific time steps with different missing ratios and investigate whether SAGE can still identify consistent governing dynamics **(Fig. 7a, c, d)**.

As shown in **Fig. 7c**, for the synthetic network, although the state reconstruction error gradually increases with increasing missing ratios, the inferred node dynamics remain highly consistent with those obtained from complete observations. Even when 50% of temporal observations are removed, SAGE can still recover the underlying dynamical structures and accurately reconstruct the observed state trajectories. The average increase in reconstruction error across the three communities remains below 0.1%. Similar results are obtained for the real-world urban rail transit network **(Fig. 7d)**. Although missing observations introduce additional uncertainty, the discovered dynamics remain stable and the average increase in reconstruction error remains below 3%. These results demonstrate that SAGE can reliably identify governing dynamics even when temporal observations are incomplete, indicating its robustness to missing measurements.

### *Transfer of governing laws to unobserved nodes*

In some real-world scenarios, certain nodes may have no observations due to sensor failures or deployment limitations. This situation creates a more challenging problem because the governing dynamics of these nodes cannot be directly identified from their own measurements. To evaluate whether SAGE can transfer learned dynamical knowledge across nodes, we completely remove observations from selected nodes and infer their dynamics using the community-level governing equations learned from observed nodes within the same community. The inferred equations are then transferred to unobserved nodes to evaluate whether shared community-level dynamical patterns can characterize nodes without direct observations **(Fig. 7b, e, f)**.

For the synthetic network **(Fig. 7e)**, the reconstructed dynamics and predicted states closely match the ground truth. The average increase in reconstruction error across the three communities remains below 0.1%. For the real-world urban rail transit network **(Fig. 7f)**, SAGE also maintains reliable reconstruction performance, with the average reconstruction error increase remaining below 15%. These results indicate that SAGE can extract shared community-level dynamical patterns from observed nodes and transfer them to nodes with completely missing observations.

### *Adaptive reasoning improves equation discovery*

To further evaluate the reliability and performance of SAGE, we conduct a series of analyses **(Fig. 8)**. These experiments examine whether SAGE can recover compact and accurate governing equations across synthetic and real-world networks and further evaluate the contributions of its main components. Increasing the relative threshold gradually reduces the number of retained terms **(Fig. 8a)**. At lower thresholds, most of the removed terms are false positives, whereas true positive terms are largely preserved and the state WMAPE remains nearly unchanged. As the threshold increases further, some true-positive terms are also removed, causing a clear increase in state WMAPE. These results indicate that a moderate threshold can remove redundant terms while preserving the terms required to accurately describe the system dynamics.

We next benchmark SAGE against several state-of-the-art equation-discovery methods, including SINDy [25], PDE-FIND [26], Two-Phase [40], LLC [22], ND2 [38] and BILLIE [32]—using both a single-dynamics synthetic network and a real-world urban rail transit network **(Fig. 8b)**. The comparison is based on WMAPE, where a lower value indicates a smaller discrepancy between the reconstructed and observed system states. SAGE consistently achieves the lowest reconstruction error on both networks, with a WMAPE of 9.0% on the real-world network and 0.60% on the synthetic network. In comparison, the best-performing baselines reach 15.8% and 0.68%, respectively. These results demonstrate that SAGE can recover governing equations that reproduce the observed system dynamics more accurately across both synthetic and real-world settings.

We compare different agent reasoning strategies, including One-shot [56], ReAct [57], Reflexion [44], Self-Refine [45], and SAGE, in terms of recovery accuracy, stability, and structural recovery rate **(Fig. 8c)**. Recovery accuracy evaluates how accurately the governing equations are recovered, stability measures the consistency of the predictive performance of the discovered equations across repeated runs, and structural recovery rate measures whether the correct functional structure is successfully identified. SAGE achieves scores of approximately 0.97, 1.00, and 1.00 for recovery accuracy, stability, and structural recovery rate, respectively. In comparison, ReAct, Reflection, and Self-Refine achieve recovery accuracies of about 0.95, but their stability remains around 0.71 and their structural recovery rates range only from approximately 0.18 to 0.27. SAGE achieves consistently high scores across all three metrics, indicating that it not only recovers the governing equations accurately, but also produces equations with more stable predictive performance and more reliable structural recovery.

Ablation experiments further validate the contribution of each module in SAGE **(Fig. 8 d-f)**. Specifically, the three ablated variants respectively remove the structure-guided candidate library (E1-library), the review process for evaluating discovered equations (E2-Review), and the feedback mechanism that uses review results to refine subsequent discovery (E3-Feedback). As shown in **Fig. 8d**, the full SAGE achieves F1 scores of approximately 1.00 and 0.97 for $dx/dt$ and $dy/dt$, respectively. Removing the structure-guided candidate library causes the largest degradation, reducing the corresponding F1 scores to approximately 0.50 and 0.45. Removing the review or feedback components also reduces structural recovery, with the F1 scores decreasing to approximately 0.60 for $dx/dt$ and 0.75 $dy/dt$. These results indicate that the candidate library is particularly important for recovering the correct functional structure, while the review and feedback mechanisms further improve term selection.

A similar pattern is observed in predictive performance **(Fig. 8e)**. SAGE achieves an $R^2$ close to 1.0 for both state dimensions, whereas the ablated variants decrease to approximately 0.85–0.88. This decline indicates that errors in the recovered equation structure directly affect the ability of the inferred equations to reproduce the observed dynamics.

The term-level analysis in **Fig. 8f** further explains the source of these differences. The ground-truth equations contain four terms for and $dx/dt$ and three terms for $dy/dt$, as indicated by the dashed reference lines. SAGE recovers exactly these four and three terms without introducing spurious terms. In contrast, removing the candidate-library component recovers only two correct terms for each equation while introducing additional false terms, resulting in four selected terms for $dx/dt$ and six for $dy/dt$. When the review or feedback mechanism is removed, the inferred $dx/dt$ equation contains two correct terms together with one spurious term, whereas the $dy/dt$ equation retains the three ground-truth terms but introduces two additional spurious terms.

Thus, the lower F1 scores of the ablated variants arise from both missed ground-truth terms and the inclusion of unnecessary functional terms. Together, these results show that the structure-guided candidate library provides an effective search space, while the review and feedback mechanisms suppress spurious terms and improve the recovered equation structure. More importantly, these findings highlight the role of LLM-guided reasoning in enabling adaptive equation discovery. By interpreting residual patterns and review feedback, the LLM identifies potential missing dynamical mechanisms and guides subsequent candidate-library updates. This adaptive refinement allows SAGE to adjust the search space according to different dynamical systems and discovery failures.

***Blind recovery of unseen dynamical mechanisms***

To determine whether adaptive candidate-space refinement provides benefits beyond simply enlarging the function library, we designed a controlled blind-repair experiment in which an entire dynamical mechanism was removed from the initial candidate library. Specifically, the state–time modulation family, including $x_i \sin(w_g t)$ , $y_i \sin(w_g t)$ , and their corresponding cross-state forms, was withheld from a synthetic community-structured network. The withheld family was removed from the initial search library but remained available in a broader candidate-family catalog that all adaptive repair methods could access. The ground-truth equation structure was never exposed during the search.

We compared four candidate-space update strategies under otherwise identical conditions. The fixed-library variant retained the incomplete initial library throughout the experiment and did not allow any new function families to be added. Random expansion added one previously unused function family, sampled uniformly from the same candidate family catalog, whenever the current equation failed the same predefined review criteria. Frozen rule-based repair used a fixed set of if–then rules that mapped residual patterns, such as periodicity or temporal drift, to candidate-function families. These rules were predefined, were not tailored to the withheld mechanism, and remained unchanged throughout the experiment. SAGE instead selected repair families from the same catalog by interpreting numerical residual patterns together with structured feedback from the Review Agent. All four methods used the same Ridge-STLSQ solver, training/validation split, candidate-family catalog, pruning procedure, and a maximum of three repair rounds. At most one new function family was introduced in each repair round.

Across 20 repeated runs, SAGE reintroduced the withheld state–time modulation family in all runs and exactly recovered the complete ground-truth equation structure in 90% of runs. Here, exact recovery means that all ground-truth terms were recovered without introducing any spurious terms. Random expansion achieved exact recovery in 50% of runs, whereas neither the fixed-library nor the frozen rule-based variant recovered the complete structure **(Fig. 8g)**. The exact recovery rate of SAGE was significantly higher than that of random expansion (one-sided Fisher's exact test, $p=0.0069$). SAGE further achieved a term-level F1 score of 0.993 and retained only 0.10 false-positive terms on average **(Fig. 8 h, i)**. These results indicate that SAGE does not benefit merely from access to a larger candidate space; rather, it uses residual and review evidence to identify which dynamical family is missing and to repair the search space selectively.

## DISCUSSION

This study introduces SAGE as an interpretable framework for the data-driven discovery of governing laws in community-structured complex networks. Rather than treating all nodes as governed by a single universal equation or fitting each node independently, SAGE addresses a central scientific question: which dynamical mechanisms are shared within a community, and which arise from individual-specific variations. A central feature of SAGE is its hierarchical formulation of network dynamics, which separates the governing dynamics of each node into community-level dynamics shared by nodes within the same community and individual-level dynamics specific to individual nodes. This decomposition provides an explicit mathematical representation of hierarchical organization in network dynamics and makes the discovered mechanisms directly interpretable. SAGE organizes governing-equation discovery into an iterative Library–Discovery–Review loop, in which the candidate function space is progressively revised according to residual patterns and review evidence rather than remaining fixed throughout the search. The resulting discovery process is therefore not only interpretable and traceable, but also able to revise its search space when the current equation fails to explain the observed dynamics.

Experiments on two synthetic and four real-world community-structured complex networks demonstrate the effectiveness of SAGE across different dynamical systems and application domains. On synthetic networks, SAGE accurately recovers the underlying governing equations and distinguishes community-level dynamical terms from individual-level dynamical terms, even when different communities follow different nonlinear dynamical systems. More importantly, the real-world experiments show that the discovered equations reveal system-specific dynamical laws rather than merely reproducing observed trajectories. In the urban mobile communication network, the equations reveal recurrent state–time modulation and nonlinear state dependence; in the EEG network, they identify cross-variable coupling underlying oscillatory dynamics; in the power grid, they recover topology-dependent phase coupling; and in the urban rail transit network, they reveal temporal dependence, nonlinear response, and passenger-flow interactions. Across these networks, the hierarchical equations further distinguish mechanisms shared by related nodes from node-specific dynamical deviations, providing an interpretable description of both collective and individual behavior. Node-state reconstruction and prediction serve as quantitative validation that these discovered laws reproduce the observed network dynamics, rather than as the primary objective of the framework. Under incomplete observations, community-level dynamics learned from observed nodes can also be transferred to nodes with completely missing observations, demonstrating the value of explicitly modeling shared dynamical patterns within communities.

The blind repair experiment provides additional evidence that adaptive candidate-space refinement is not equivalent to simply enlarging the function library. When the entire state–time modulation family was withheld from the initial library, SAGE recovered the missing family in all 20 repeated runs and exactly recovered the complete governing structure in 90% of runs, compared with 50% for random expansion and 0% for both the fixed-library and frozen rule-based variants. SAGE also achieved a term-level F1 score of 0.993 while retaining only 0.10 false-positive terms on average. These results support the role of evidence-guided reasoning in identifying which dynamical mechanism is missing and repairing the candidate space in a targeted manner. Comparisons with representative equation-discovery methods, different agent reasoning strategies, and ablation experiments further demonstrate the advantages of SAGE in structural recovery, predictive performance, and solution stability.

Beyond predictive performance, the main methodological significance of SAGE lies in combining numerical equation discovery with LLM-guided scientific reasoning while preserving explicit mathematical laws. Most sparse equation-discovery methods rely on a predefined candidate function library, so relevant dynamical terms may be missed when the required functions are absent from the initial library. SAGE instead treats the candidate function library as an adaptive space that can be revised according to residual patterns and review evidence. Importantly, the LLM does not directly generate governing equations or estimate numerical coefficients. Numerical methods remain responsible for generating candidate equation structures and estimating their coefficients, while the LLM uses structured diagnostic evidence to guide candidate selection and candidate-library refinement. This separation is important for interpretability: the numerical solver determines the equation and its coefficients, whereas the LLM helps decide where the search space should be revised. Because each retained term has an explicit mathematical form and can be linked to a corresponding dynamical mechanism, the discovered equations remain physically interpretable rather than functioning as black-box predictors. Together, hierarchical decomposition and adaptive candidate-space refinement turn equation discovery into a process for identifying shared and individual dynamical laws from observational network data.

### *Limitations of the study*

Nevertheless, several limitations remain. First, the separation between community-level and individual-level dynamics relies on the quality of community identification. In this study, community structures are either obtained from available domain knowledge or inferred from node-state observations when prior labels are unavailable. While this enables application to networks without predefined communities, inaccurate or ambiguous community assignments may affect the decomposition of shared and individual dynamics. Future work could jointly optimize community identification and governing-equation discovery, allowing the discovered dynamics to further refine the community assignments. Second, strong observational noise, irregular sampling, or sparse measurements may affect derivative estimation and subsequent equation-structure identification. More robust derivative-estimation techniques or derivative-free equation-discovery methods could therefore be incorporated. Third, although SAGE can adaptively revise the candidate library, the current implementation selects new function families from a predefined catalog. The blind repair experiment therefore tests the recovery of mechanisms withheld from the initial library rather than completely unrestricted generation of previously unknown mathematical forms. Extending the framework toward more open-ended function discovery is an important direction for future work. Finally, the current formulation assumes that community assignments and governing equations remain approximately stable over the observation period. Extending SAGE to identify time-varying community structures and evolving governing equations would further broaden its applicability to real-world complex networks.

## METHODS

### *Library Agent*

The Library Agent prepares the observational data and manages the candidate function library throughout equation discovery. It first preprocesses node-state observations and determines community assignments using either available domain knowledge or inferred dynamical structures from node-state observations when prior information is unavailable. The community assignments obtained are then used to organize nodes for subsequent hierarchical equation discovery. It then constructs an initial candidate function library from the processed observations. During subsequent search rounds, the Library Agent updates the library in response to residual patterns and diagnostic feedback from the Review Agent, adding function families that may capture missing dynamical mechanisms.

Using the preprocessed node-state observations and, when available, the adjacency matrix, the Library Agent constructs the initial candidate function library. The candidate functions are evaluated for each node using the same library and are then organized according to the community assignments. For community k, resulting initial candidate-function matrix is written as:

$$\boldsymbol{\Theta}_K^{(0)} = [\boldsymbol{\Theta}^{(0)}(\boldsymbol{x}_i(t))]_{i \in k} \tag{10}$$

where $\boldsymbol{\Theta}^{(0)}(\boldsymbol{x}_i(t))$ denotes the candidate function vector generated from the initial library and evaluated at the state of node $i$ at time $t$, and $k$ indicates the community index. When community labels are unavailable, they are inferred from the node-state observations during preprocessing (Community Clustering). Thus, all nodes share the same candidate function definition, whereas the community labels determine how their observations are assembled for community-level equation discovery. The complete set of candidate function families is provided in Table S2.

The library is used in the first discovery round but may not contain all functions required to explain the observed dynamics. After the Discovery Agent selects an equation at search round $r$, the Review Agent evaluates it on the validation set. The corresponding derivative residual is defined as:

$$\mathrm{Res}^{(r)} = \frac{\mathrm{d}x_i(t)}{\mathrm{d}t} - \frac{\mathrm{d}\hat{x}_i^{(r)}(t)}{\mathrm{d}t} \tag{11}$$

where $\frac{\mathrm{d}x_i(t)}{\mathrm{d}t} \in \mathbb{R}^d$ denotes the derivatives estimated from the validation data, and $\frac{\mathrm{d}\hat{x}_i^{(r)}(t)}{\mathrm{d}t} \in \mathbb{R}^d$ denotes the derivatives predicted by the equation identified at search round. The Review Agent summarizes the failed review checks, residual patterns, and unstable terms as diagnostic feedback. When this feedback suggests that the current library lacks relevant dynamical mechanisms, the LLM identifies additional or revised function families from the broader structural catalog and uses them to generate or instantiate candidate terms that are not currently represented in the active library. Let $R^{(r)}$ denote the function families selected for repair at search round r. The candidate function library is updated as:

$$L^{(r+1)} = P[L^{(r)} \cup S(REPAIR^{(r)})] \tag{12}$$

where $S(REPAIR^{(r)})$ maps the selected function families to candidate terms, and $P(\cdot)$ denotes the library pruning operator that removes duplicate and strongly collinear terms. The detailed procedure for function family selection, candidate term generation, and library pruning is described in Document S1, section 1.3. If a selected function family is already represented in the current library, the framework may instantiate higher-order, reparametrized, or interaction variants constructed from the admissible primitives rather than simply duplicating existing terms. The updated library is then returned to the Discovery Agent for the next round of equation structure search, completing the Library–Discovery–Review repair loop. Implementation details of the LLM interface are provided in Document S1, section 1.6.

### *Discovery Agent*

The Discovery Agent searches the current candidate library for sparse governing equations. The Discovery Agent is not tied to any specific numerical solver. Numerical methods are used to generate candidate equation structures and estimate their coefficients and other suitable methods can be incorporated into the framework. In this study, we use bootstraps sequentially thresholded least squares (STLSQ) [24], Lasso with cross-validation [58], and Ridge regression [59] followed by validation-guided thresholding as three representative methods. These methods employ different regularization and term-selection mechanisms and therefore provide complementary candidate structures and supporting evidence for reliable term selection. The three methods used here are not fixed components of SAGE and can be replaced or extended by other numerical equation-discovery methods without changing the overall framework. At the community level, each numerical backend estimates the shared coefficient vector $\hat{\boldsymbol{\xi}}_k^S$ defined in the governing equation formulation above using observations from nodes within community $k$.

The numerical backends return candidate equation structures and their estimated coefficients. For each candidate term $m$, the Discovery Agent evaluates three complementary stability measures: $f_m$ , the retention consistency across numerical backends; $b_m$, the selection frequency across bootstrap samples; and $s_m$, the consistency of the coefficient signs across repeated estimates. Together, these measures define the stability evidence for term m as:

$$z_m = (f_m, b_m, s_m) \tag{13}$$

These stability measures are retained separately and provide term-level supporting evidence for each candidate equation; no single stability measure determines whether a term is finally selected. Each candidate equation is then evaluated by its validation error and structural complexity, where structural complexity is measured by the number of retained terms. The Pareto candidate set is defined as:

$$\mathcal{P}_{\text{Pareto}} = \{q \in \mathcal{Q} | \nexists q' \in \mathcal{Q}: E_{q'} \leq E_q,\ C_{q'} \leq C_q,\ (E_{q'} < E_q \vee C_{q'} < C_q)\} \tag{14}$$

Here, $E_q$ and $C_q$ denote the validation error and structural complexity of candidate equation, respectively. Pareto filtering removes candidates that are both less accurate and more complex, while retaining alternative structures that represent different trade-offs between accuracy and simplicity. The LLM then compares only these Pareto candidates using their validation error, structural complexity, and the stability evidence of their retained terms, and selects the community-level equation structure. The LLM does not generate new equation structures or estimate coefficients. After the structure is selected, the equation structure is fixed and its coefficients are re-estimated numerically using observations from nodes in the corresponding community. This refitting step follows the same interchangeable numerical-backend interface and is therefore not tied to a particular solver.

The Discovery Agent next examines the part of each node's dynamics that remains unexplained by the selected community-level equation. For node i in community k, the residual dynamics are obtained by subtracting the fitted community-level component from its observed temporal derivative:

$$Res_i^H(t) = \frac{dx_i(t)}{dt} - \Theta^{(r)}(x_i(t))\boldsymbol{\xi}_k^{S(r)} \tag{15}$$

The residual dynamics are then examined to determine whether additional individual-level terms are required. A candidate term is treated as an individual-level terms if it improves validation performance for only a subset of nodes, or if it is consistently selected across nodes within the same community but produces different fitted coefficient values between nodes. These two forms of evidence allow the Discovery Agent to distinguish individual-level variations from the dynamics already explained by the shared community-level equation. To quantify the contribution of candidate term m to node i, we measure the reduction in validation error after including the term:

$$\Delta E_{i,m} = E_i^{(-m)} - E_i^{(+m)} \tag{16}$$

The validation improvement of candidate term for node is measured by $\Delta E_{i,m}$, defined as the reduction in validation error after including the term. For each candidate individual-level term, the Discovery Agent considers three factors: whether the term improves validation performance, whether this improvement is consistently observed across repeated fits, and how much additional complexity the term adds to the equation. The decision to retain a term is based on these factors together rather than on any single criterion. After the community-level and individual-level terms are determined, their coefficients are refitted numerically using the selected numerical backend. Finally, any term whose removal does not noticeably reduce validation performance is discarded. The resulting governing equation follows the same formulation as introduced in Equation 2.

### *Review Agent*

The Review Agent evaluates each selected equation on the validation set using three complementary checks: derivative fit, coefficient stability, and state prediction. Derivative fit measures the agreement between the derivatives predicted by the inferred equation and the observed derivatives. To assess coefficient stability, the selected structure is fixed and its coefficients are re-estimated over B bootstrap resamples. For candidate term m, the relative coefficient dispersion is defined as:

$$CV_m = \frac{std(\xi_m^{(1)}, \dots, \xi_m^{(B)})}{|mean(\xi_m^{(1)}, \dots, \xi_m^{(B)})|} \tag{17}$$

where a smaller $CV_m$ indicates a more stable coefficient estimate. State prediction further evaluates whether the inferred equation can accurately predict future system states by evolving the observed initial states and comparing the predicted trajectories with the actual validation trajectories. Using the three checks jointly, the Review Agent returns ACCEPT when the equation satisfies the predefined review criteria, REPAIR when the remaining errors indicate a correctable structural mismatch and REJECT when the available data cannot support a stable equation or the maximum number of repair rounds is reached. The detailed review criteria and numerical settings are provided in Document S1, section.

An ACCEPT decision indicates that the selected equation satisfies the predefined validation criteria and represents a physically interpretable governing equation, with each retained functional term corresponding to an identifiable dynamical mechanism. The accepted equation is then passed to the independent test set for final evaluation. For REPAIR, the Review Agent summarizes the failed checks, residual patterns, and unstable terms into a structured repair plan. Based on this evidence, the LLM selects or refines function families and proposes admissible candidate-term constructions from the available mathematical primitives, while numerical solvers remain solely responsible for coefficient estimation and final equation fitting. Let $REPAIR^{(r)}$ denote the function families selected for repair at round r. The resulting repair instruction is passed to the Library Agent, which updates the candidate function library according to Equation 3. The updated library is then passed back to the Discovery Agent, and the Library–Discovery–Review cycle is repeated.

A REJECT decision terminates the search when the available data cannot support a stable equation or when the maximum number of repair rounds is reached. In this case, the system generates a structured report summarizing the entire iteration history, including the candidate library evolution, tested equations, review outcomes, and repair actions, ensuring full traceability and diagnostic transparency. Throughout the search, each round records the candidate library, candidate equations, review results, and repair actions. The test set remains isolated throughout library adaptation, model selection, and review, and is used only after an equation receives ACCEPT.

### *Community Clustering*

When community labels are provided, they are directly adopted as prior structural information. When labels are unavailable, the Library Agent infers dynamical communities from node-state observations using fuzzy K-means clustering. This state-based approach assigns nodes with similar dynamical behaviors to the same community and does not require predefined network topology information for community identification.

### *Numerical Differentiation*

Accurate derivative estimation is critical for equation discovery but remains challenging for noisy observational data. We therefore use the Savitzky–Golay filter [60], which applies local polynomial regression to smooth the data while preserving higher order moments. This approach suppresses noise and provides stable derivative estimates.

### *Evaluation Metric*

To quantify errors in the inferred node dynamics and predicted node states, we use the weighted mean absolute percentage error (WMAPE):

$$\mathrm{WMAPE} = \frac{\sum_{i=1}^{N} |y_i - \hat{y}_i|}{\sum_{i=1}^{N} |y_i|} \times 100\% \tag{18}$$

Here, $y_i$ and $\hat{y}_i$ denote the true and predicted values, respectively, and is the number of samples.

### *Ethics statement*

This study involved no new experiments on human participants or animals. Therefore, ethical approval and informed consent were not required.

## RESOURCE AVAILABILITY

### *Lead contact*

Requests for further information and resources should be directed to and will be fulfilled by the lead contact, Jinlei Zhang (zhangjinlei@bjtu.edu.cn).

### *Materials availability*

This study did not generate new physical materials.

### *Data and code availability*

The synthetic and real-world datasets used in this study are publicly available in the SAGEdata repository on Hugging Face: https://huggingface.co/datasets/tczzx6/SAGEdata/tree/main. The repository includes the synthetic single- and multi-dynamics datasets and the real-world urban mobile communication, brain EEG, IEEE 39-bus power grid, and Beijing urban rail transit datasets. Processed data and data-generation scripts required to reproduce the main analyses are also provided. Further details are given in the Methods and Document S1.

The implementation of SAGE is publicly available at https://huggingface.co/tczzx6/LLM-SAGE/tree/main. The repository includes the Library Agent, Discovery Agent, and Review Agent, together with the numerical equation-discovery modules, adaptive candidate-library refinement procedure, and scripts used to reproduce the main experiments and figures reported in this study.

Any additional information required to reanalyze the data reported in this paper is available from the lead contact upon request.

## ACKNOWLEDGMENTS

This work was supported in part by the National Natural Science Foundation of China under Grant 72288101, in part by the Beijing Natural Science Foundation under Grant 4252049 and Grant L251030, and in part by the Fundamental and Interdisciplinary Disciplines Breakthrough Plan of the Ministry of Education of China under Grant JYB2025XDXM211.

The authors have no additional acknowledgements to declare.

## AUTHOR CONTRIBUTIONS

J.Z. conceived the study and developed the overall research framework. J.Z., Z.Z., and H.Q. developed the methodology and performed the main analyses. Z.Z. and H.Q. conducted the experiments and data analysis. F.C., J.Y., X.Y., X.M., H.S., and J.G. contributed to the result interpretation and discussion. L.Y. and Z.G. supervised the research. J.Z., Z.Z., and H.Q. wrote the initial manuscript. All authors reviewed and revised the manuscript and approved the final version.

## DECLARATION OF INTERESTS

The authors declare no competing interests.

## SUPPLEMENTAL INFORMATION

Document S1. Supplemental methods, Tables S1–S6, and Figures S1–S6.

## FIGURE TITLES AND LEGENDS

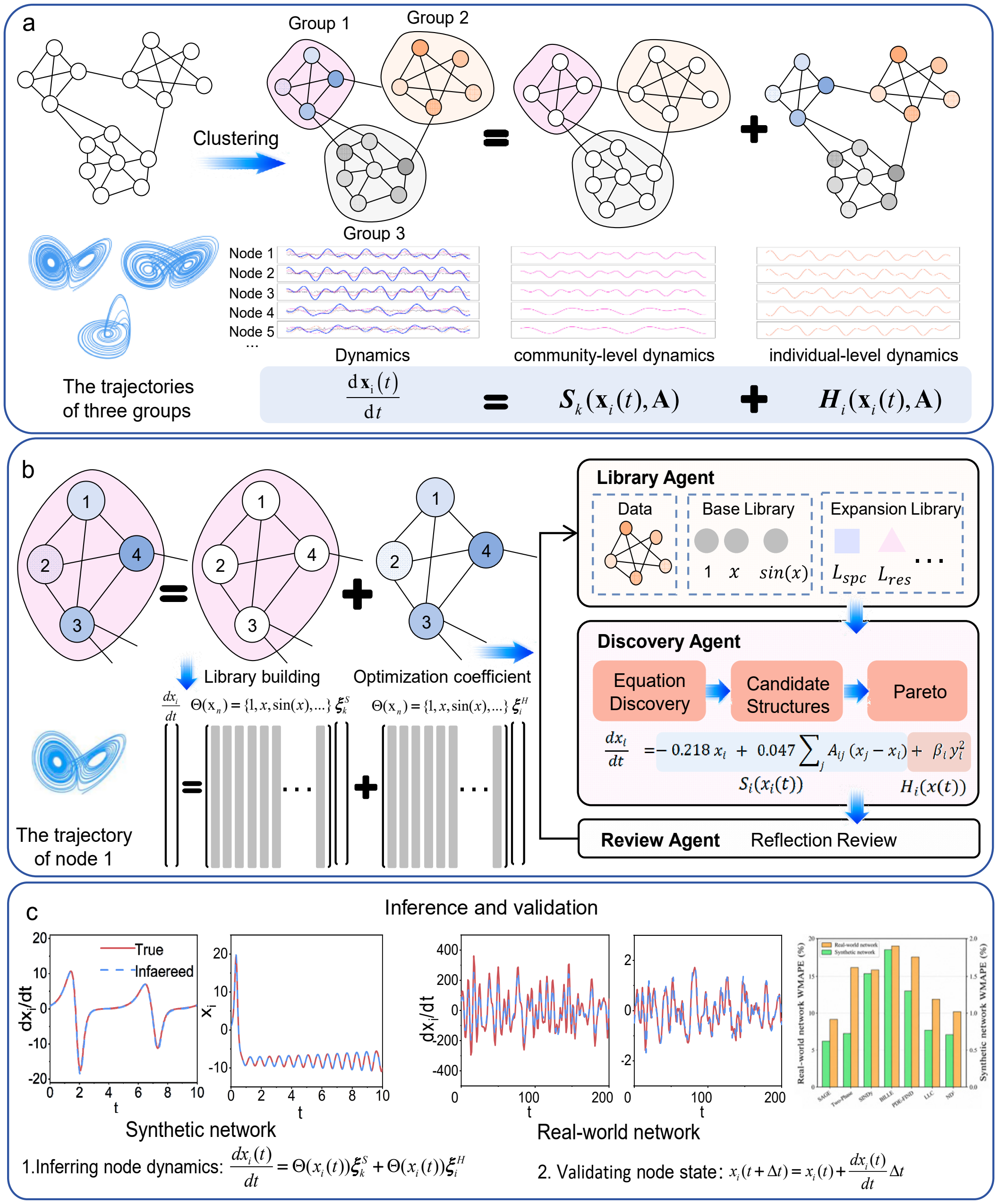

**Figure 1. Overview of SAGE. a)** Nodes are organized into dynamical communities, where community assignments can be obtained from prior knowledge when available or inferred from node-state observations when such information is unavailable. The community-level dynamics shared within each community are then modeled separately from the individual-level dynamics specific to each node. **b)** Dynamics inference for a representative node from the first community. The Library Agent constructs and adaptively updates the candidate function library from observational data. The Discovery Agent generates candidate equation structures and selects Pareto-optimal solutions. The Review Agent evaluates the selected equations and provides feedback for subsequent library refinement. **c)** Inferred node dynamics and predicted node states for synthetic and real-world networks, together with comparisons against baseline method.

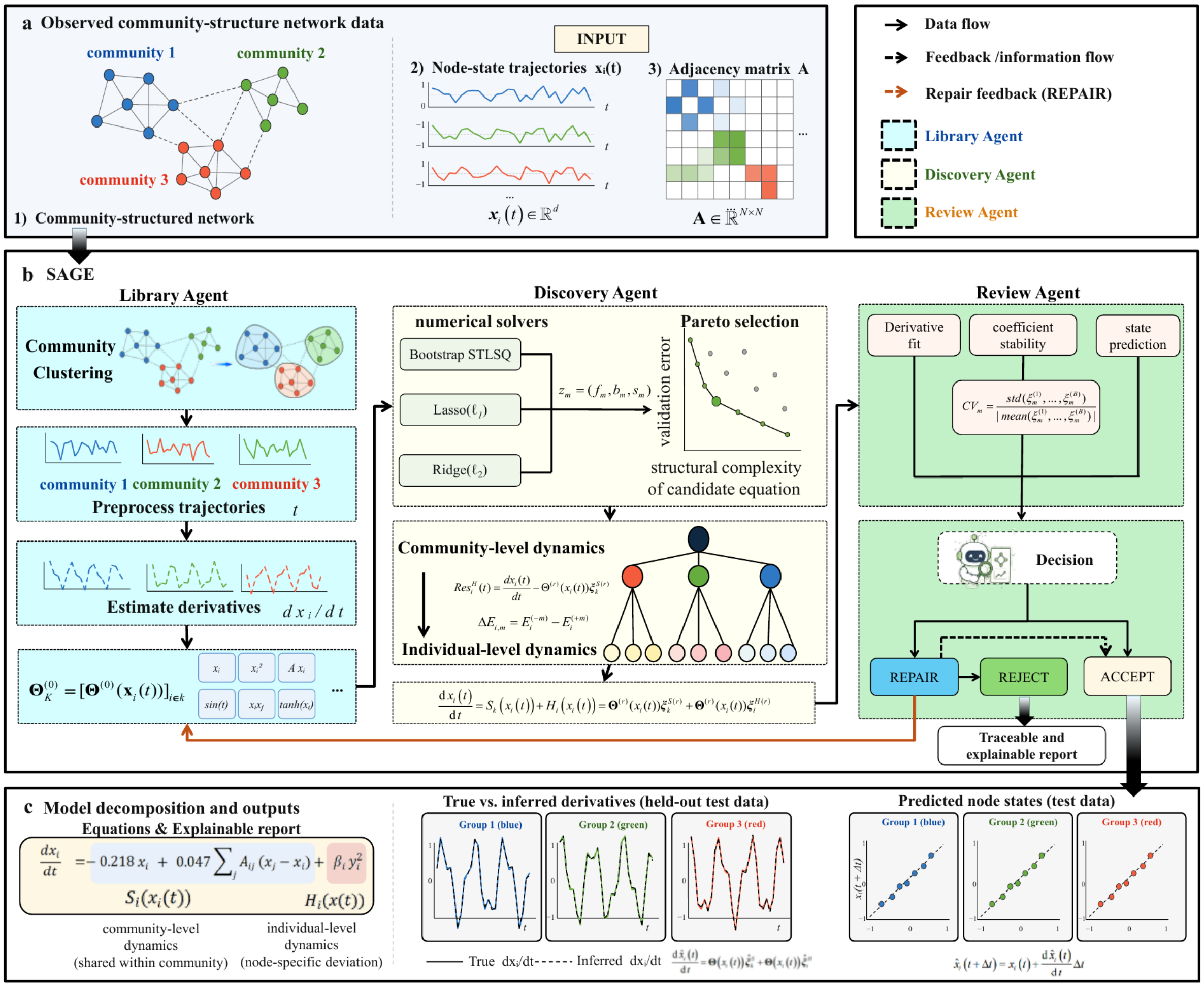

**Figure 2. Overview of the SAGE framework for governing-equation discovery in community-structured complex networks. a)** Input data, including the community-structured network, node-state trajectories, and adjacency matrix. **b)** The iterative Library–Discovery–Review workflow. The Library Agent identifies community structures, preprocesses node-state trajectories, estimates temporal derivatives, and constructs the candidate function library. The Discovery Agent uses multiple numerical solvers to generate candidate equation structures, performs Pareto selection based on validation error and structural complexity, and separately identifies community-level and individual-level dynamics. The Review Agent evaluates the selected equations using derivative fit, coefficient stability, and state prediction, and returns ACCEPT, REPAIR, or REJECT. REPAIR feedback is returned to the Library Agent to guide subsequent library refinement and equation discovery. **c)** Model decomposition and final outputs, including the inferred community-level and individual-level governing terms, true-versus-inferred derivatives on held-out test data, predicted node states on the test set, and a traceable and explainable report.

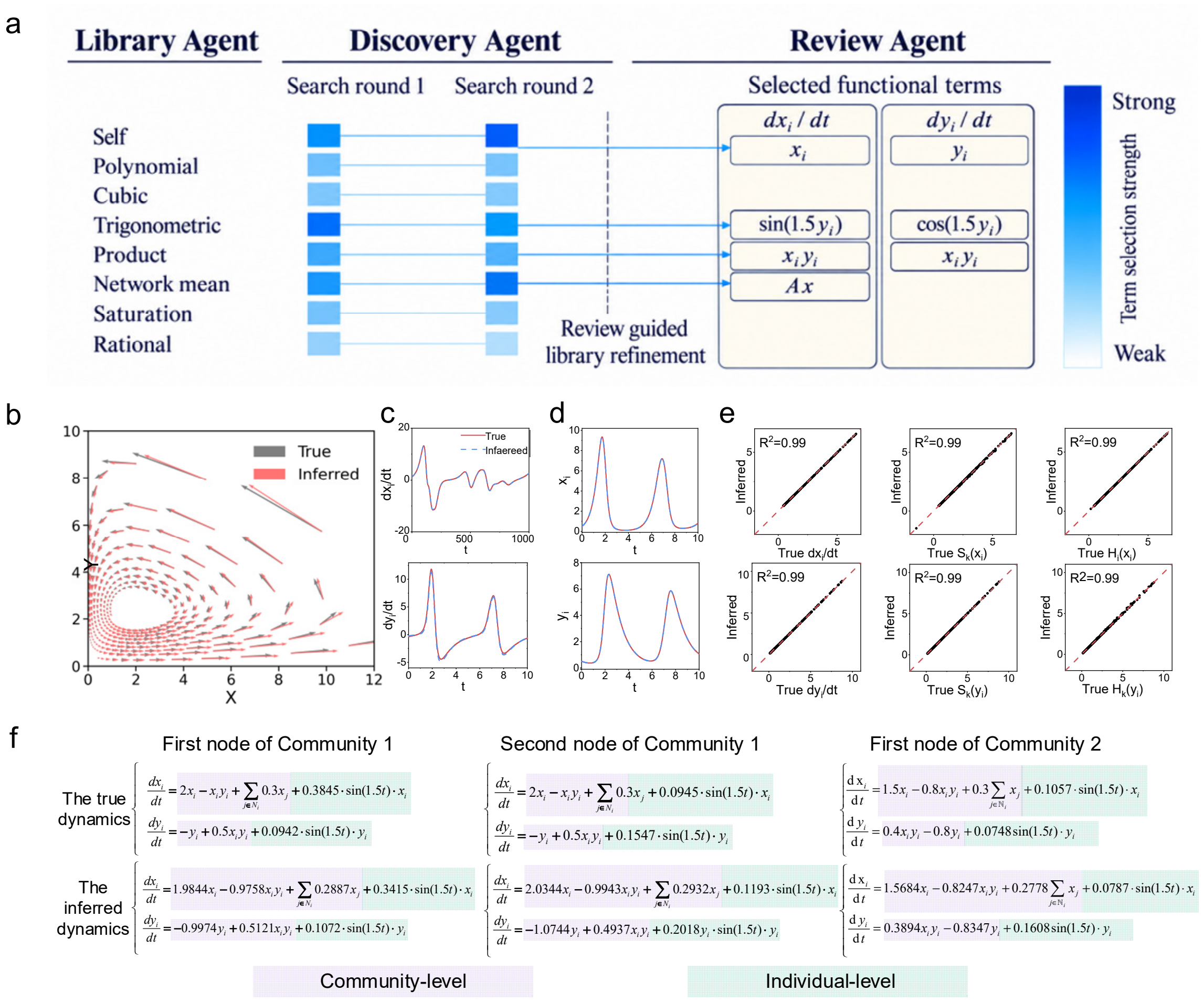

**Figure 3. Validating SAGE on a single-dynamics network with 60 nodes divided into three communities. a)** Evolution of candidate terms guided by the Library–Discovery–Review Agent loop. Candidate terms are progressively selected from the function library based on their contribution to equation discovery, with color intensity indicating selection strength. The inference results for a representative node in the first community. **b)** True and inferred trajectories and force field in both *x* and *y* dimensions. **c-d)** True and inferred node dynamics ($dx_i/dt$,$dy_i/dt$) and the validation using node state ($x_i$,$y_i$) for a representative node in the first community. **e)** Goodness of fit between true and inferred overall ($dx_i/dt$, $dy_i/dt$), community-level ($S_k(x_i)$, $S_k(y_i)$), and individual-level ($H_i(x_i)$, $H_i(y_i)$) dynamics. **f)** True and inferred governing equations for the first and second nodes in Community 1 and the first node in Community 2.

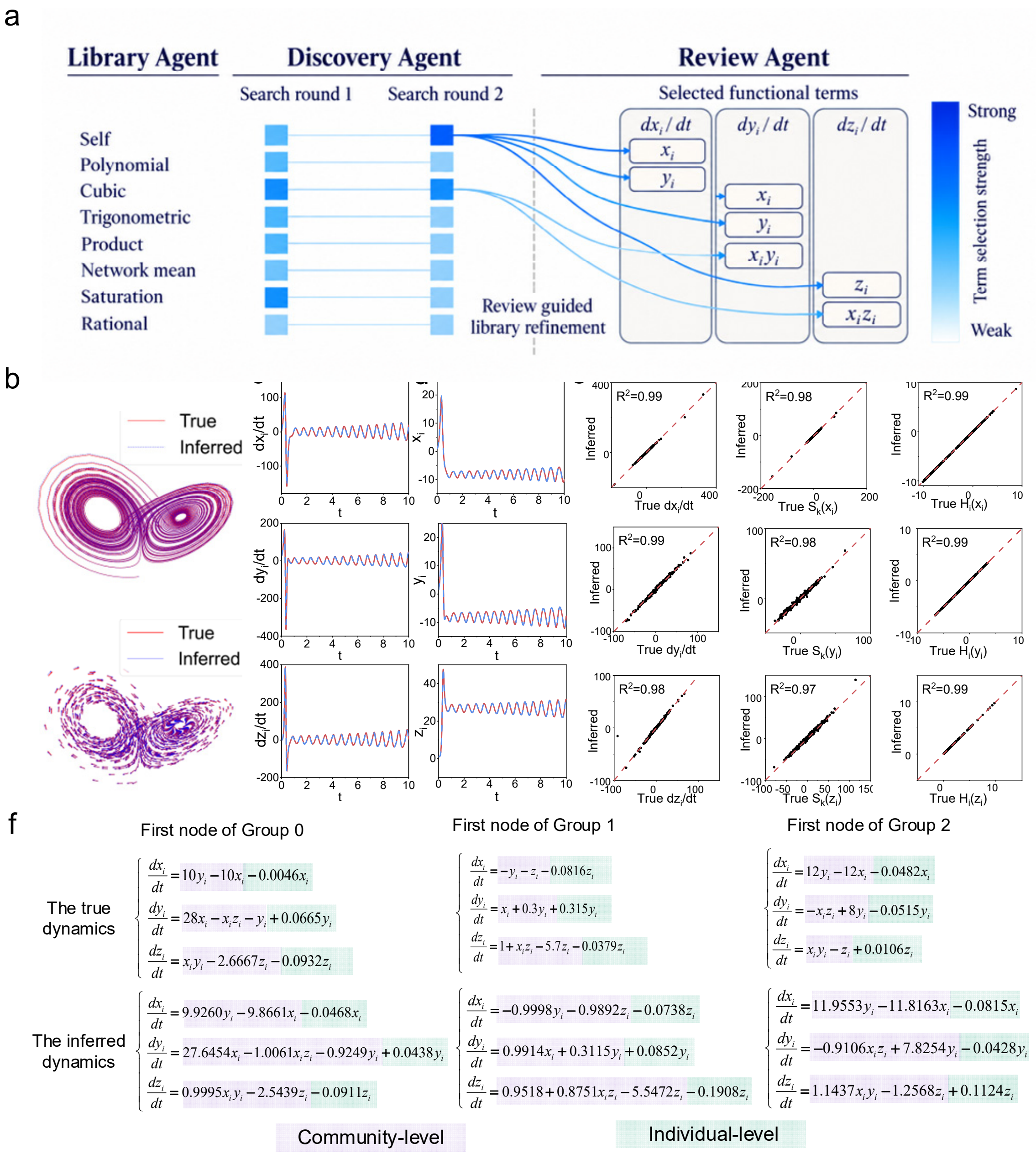

**Figure 4. Validating SAGE on a multi-dynamics network with 60 nodes divided into three communities. a)** Evolution of candidate terms guided by the Library–Discovery–Review Agent loop. Candidate terms are progressively selected from the function library based on their contribution to equation discovery, with color intensity indicating selection strength. **b)** True and inferred trajectories and force field in both x, y, and z dimensions. **c-d)** Node dynamics ($dx_i/dt$,$dy_i/dt$,$dz_i/dt$) and node state ($x_i$,$y_i$,$z_i$ ) generated by true and inferred governing equations. **e)** Goodness of fit between true and inferred overall ($dx_i/dt$,$dy_i/dt$,$dz_i/dt$), community-level ($S_k(x_i)$,$S_k(y_i)$,$S_k(z_i)$) and individual-level ($H_i(x_i)$,$H_i(y_i)$,$H_i(z_i)$) dynamics. **f)** True and inferred governing equations for the first and second nodes in Community 1 and the first node in Community 2

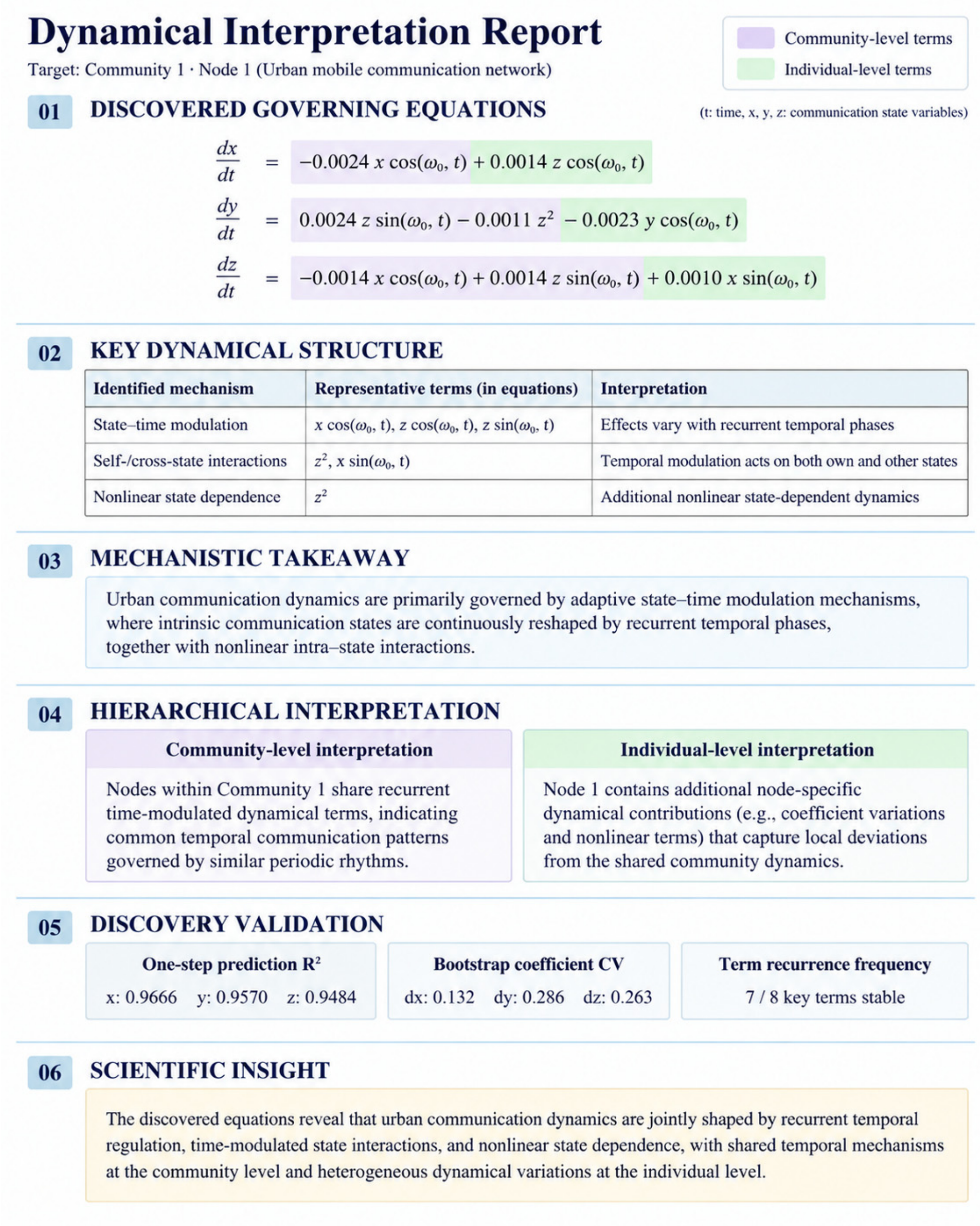

**Figure 5. Validation of governing equations discovered across real-world networks.** Report presents the discovered governing equations, key dynamical structures, mechanistic interpretation, hierarchical decomposition into community-shared and individual-specific dynamics, and quantitative validation based on step prediction $R^2$ bootstrap coefficient variation and term recurrence frequency. Purple and green shading indicate community-level and individual-level terms, respectively. The results reveal a hierarchical organization in which shared dynamical mechanisms coexist with heterogeneous node-specific variations across real-world systems.

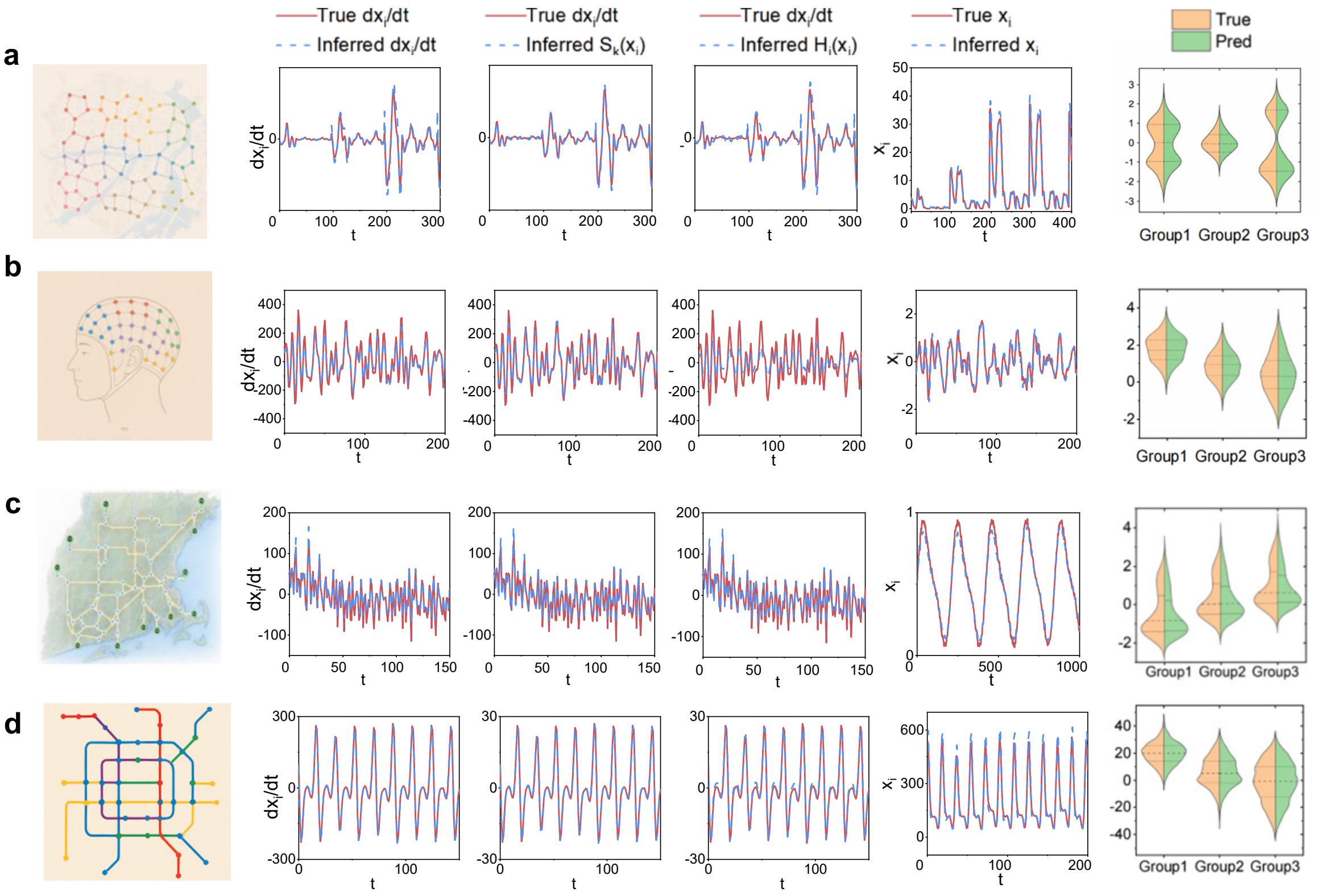

**Figure 6. Validation of SAGE on real-world networks through node-state reconstruction and comparison between observed and predicted state distributions across all nodes and communities.** **a)** Urban mobile communication network. **b)** Brain EEG network. **c)** Power grid network. **d)** Urban rail transit network.

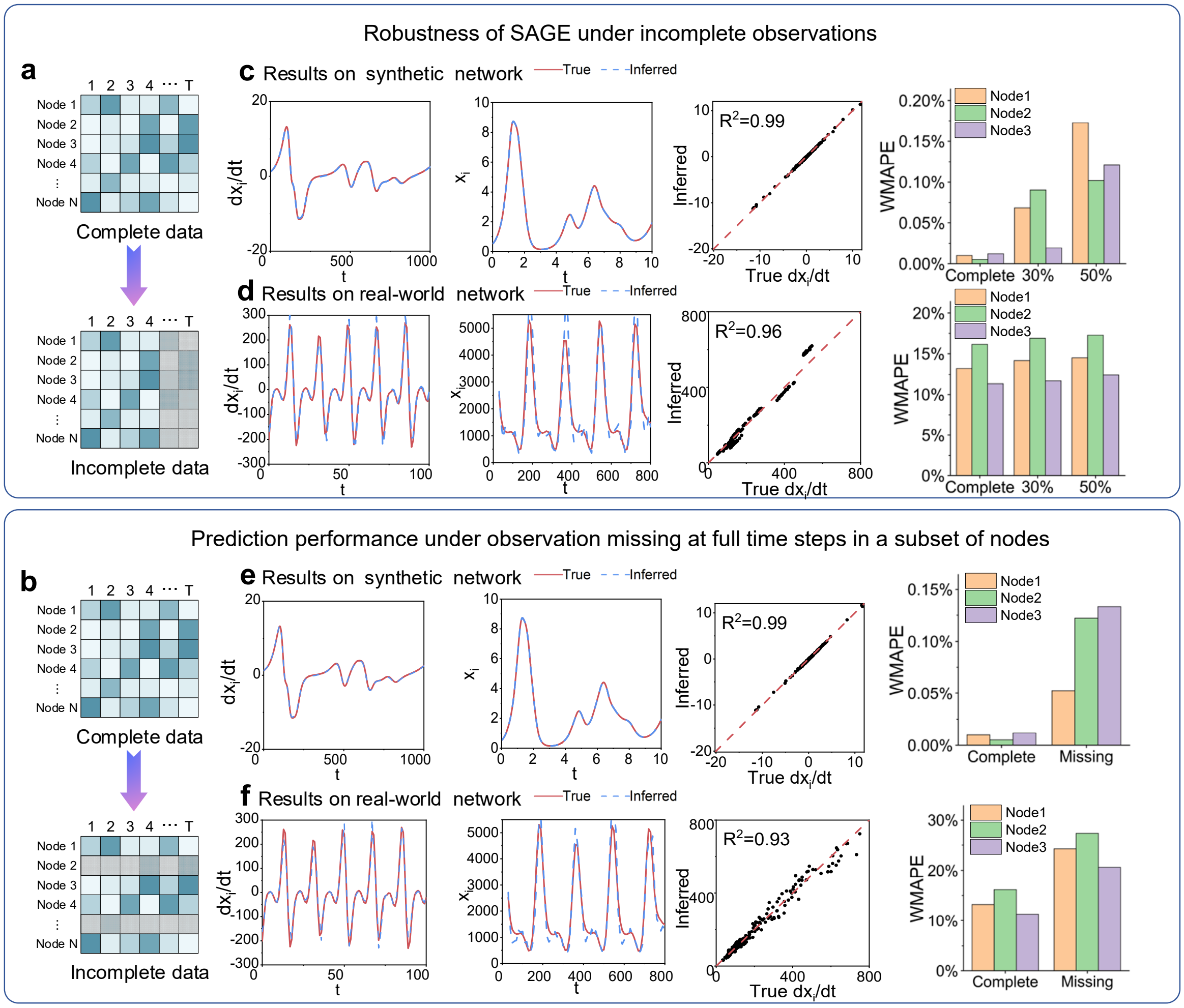

**Figure 7. Robustness of SAGE under incomplete observations. a)** Diagram of observation missing at partial time steps across all nodes. **b)** Diagram of observation missing at full-time steps in a subset of nodes. **c-f)** from left to right: true and inferred node dynamics ($dx_i/dt$) and node states ($x_i$), goodness of fit between true and inferred node dynamics ($dx_i/dt$), and performance comparison under different data missing settings for both synthetic networks **(c, e)** and real-world network **(d, f)**.

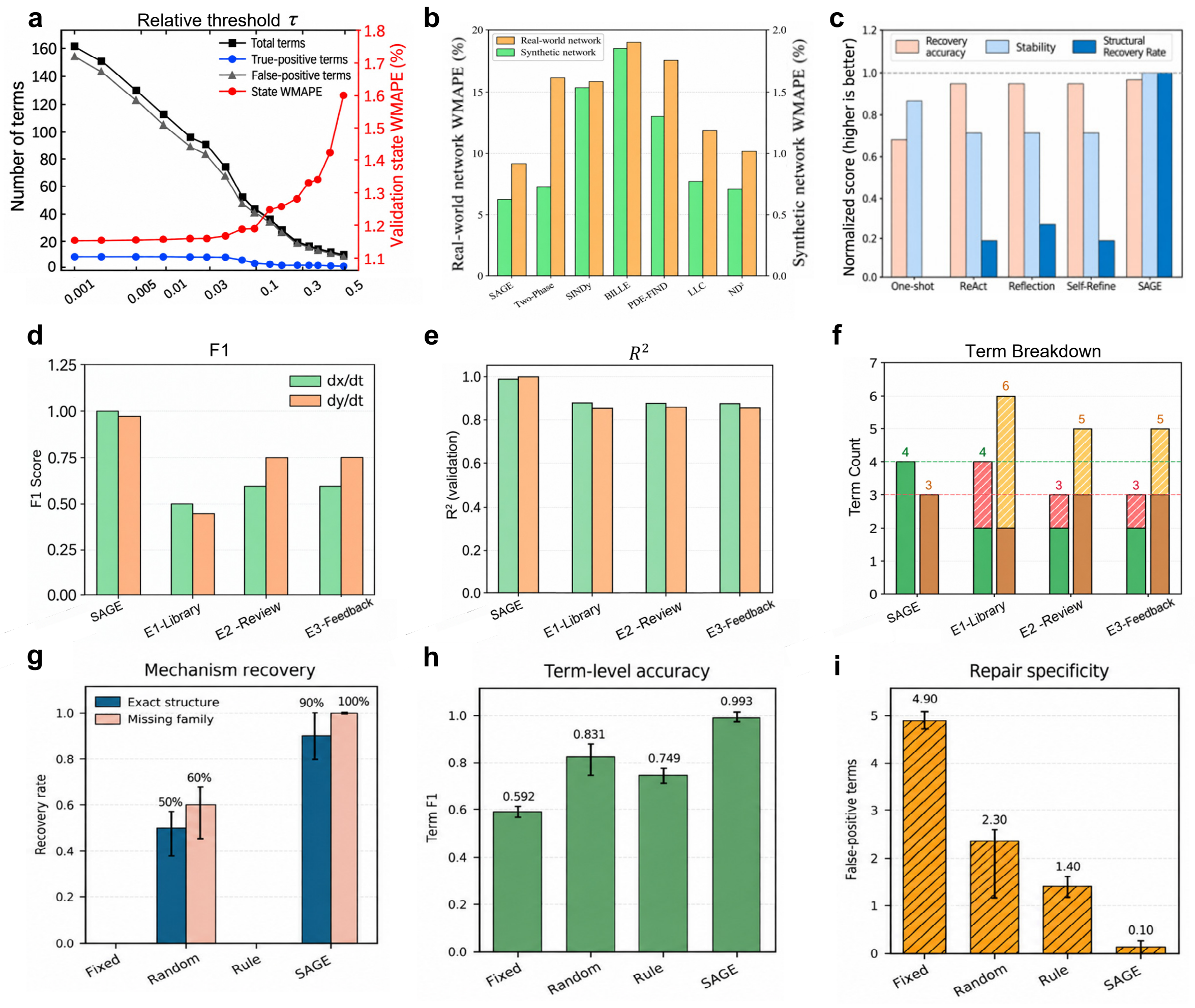

**Figure 8. Diagnostic and ablation analysis of SAGE. a)** Sensitivity to the relative threshold $\tau$ showing the numbers of total, true positive, and false positive terms together with validation state WMAPE. **b)** Reconstruction errors of SAGE and representative equation discovery methods on synthetic and real-world networks. All methods are constrained to use the same number of terms in the discovered equations. **c)** Comparison of different agent reasoning strategies in terms of recovery accuracy, stability (1-SD ($R^2$)), and structural recovery rate. **d)** Equation-discovery F1 scores for $dx/dt$ and $dy/dt$ under SAGE and different module ablations. **e)** Predictive accuracy of the discovered equations measured by $R^2$ **f)** Numbers of recovered terms for $dx/dt$ and $dy/dt$; hatched regions indicate spurious terms, and dashed lines denote the corresponding ground truth term counts. **g)** Recovery rates of the withheld state–time modulation family and the complete ground-truth equation structure across 20 repeated runs. The entire state–time modulation function family was removed from the initial candidate library before discovery. **h)** Term-level F1 scores of the recovered governing equations. **i)** Average number of false-positive terms remained in the final equations. Error bars indicate variability across repeated runs. All methods used the same Ridge-STLSQ numerical solver, candidate-family catalog, data split, and three-round repair budget. The frozen rule-based baseline used predefined repair rules that were not tailored to the withheld mechanism and remained unchanged throughout the experiment.

# Document S1. Supplemental information

## Discovering hierarchical governing equations for community-structured complex network dynamics with LLM agents

### 1. Methods

SAGE is a multi-agent framework for governing-equation discovery. It organizes equation discovery into a three-agent iterative loop. The Library Agent manages the candidate functions, the Discovery Agent identifies sparse equation structures, and the Review Agent evaluates the selected equations and requests repair when necessary.

#### *1.1 Notations and problem formulation*

For clarity, the main notations and their dimensions used in this paper are summarized in Table S1. Scalars are denoted by non-bold letters (e.g., $x_i(t) \in \mathbb{R}$), and vectors or matrices are denoted by bold letters (e.g., $\boldsymbol{x}_i(t) \in \mathbb{R}^d$). The detailed definitions of the symbols are given below.

Table S1. List of notations

| Symbol | Description |
|---|---|
| $N \in \mathbb{R}$ | Total number of nodes in the network |
| $K \in \mathbb{R}$ | Total number of communities in the network |
| $k \in \{1,2, \ldots, K\}$ | Index of community *k* |
| $T \in \mathbb{R}$ | Total number of time steps |
| $N_k \in \mathbb{R}$ | Total number of nodes in community *k* |
| $\boldsymbol{A} \in \mathbb{R}^{N\times N}$ | Network adjacency matrix describing interactions among *N* nodes. |
| $\boldsymbol{x}_\mathrm{i}(t) \in \mathbb{R}^d$ | The state of node *i* at time *t* across all *d* dimensions |
| $x_i(t) \in \mathbb{R}$ | The state of node *i* at time *t* for one dimension |
| $\frac{\mathrm{d}\boldsymbol{x}_\mathrm{i}(t)}{\mathrm{d}t} \in \mathbb{R}^d$ | The dynamics of node *i* at time *t* across all *d* dimensions |
| $\frac{\mathrm{d}x_i(t)}{\mathrm{d}t} \in \mathbb{R}$ | The dynamics of node *i* at time *t* for one dimension |

| Symbol | Description |
| --- | --- |
| $\boldsymbol{S}_k(\boldsymbol{x}_i(t), \boldsymbol{A}) \in \mathbb{R}^d$ | The community-level dynamics of community k across all d dimensions |
| $S_k \in \mathbb{R}$ | The community-level dynamics of community *k* for one dimension |
| $\boldsymbol{H}_i(\boldsymbol{x}_i(t), \boldsymbol{A}) \in \mathbb{R}^d$ | The individual-level dynamics of node *i* across all *d* dimensions |
| $H_i \in \mathbb{R}$ | The individual-level dynamics of node *i* for one dimension |
| $L^{(r)} = \{L_m\}_{m=1}^{M_r}$ | Adaptive candidate function library maintained at search round *r* |
| $\boldsymbol{\Theta}^{(r)}(\boldsymbol{x}_i(t)) \in \mathbb{R}^{M_r}$ | Candidate function vector evaluated from $L^{(r)}$ for node $i$ at time $t$ |
| $\boldsymbol{\Theta}_k^{(r)} \in \mathbb{R}^{N_k T \times M_r}$ | Candidate function matrix constructed for community *k* at round *r* |
| $M_r$ | Number of candidate functions at search round *r* |
| $\boldsymbol{\xi}_k^{S(r)} = \left[\xi_{k,1}^{S(r)}, \cdots, \xi_{k,M}^{S(r)}\right]^T \in \mathbb{R}^{M_r}$ | The coefficient vector for community-level dynamics of community *k* |
| $\boldsymbol{\xi}_i^{H(r)} = \left[\xi_{i,1}^{H(r)}, \cdots, \xi_{i,M}^{H(r)}\right]^T \in \mathbb{R}^{M_r}$ | The coefficient vector for individual-level dynamics of node *i* |
| $\boldsymbol{\xi}^{S(r)} = \left[\boldsymbol{\xi}_1^{S(r)}, \cdots, \boldsymbol{\xi}_K^{S(r)}\right] \in \mathbb{R}^{M_r \times K}$ | The coefficients for community-level dynamics of all *k* communities |
| $\boldsymbol{\xi}^{H(r)} = \left[\boldsymbol{\xi}_1^{H(r)}, \cdots, \boldsymbol{\xi}_N^{H(r)}\right] \in \mathbb{R}^{M_r \times N}$ | The coefficients for individual-level dynamics of all *N* nodes |
| $\frac{\mathrm{d}\hat{x}_i(t)}{\mathrm{d}t} \in \mathbb{R}$ | The inferred dynamics of node *i* at time *t* |
| $\Delta t$ | Prediction time interval |
| $\hat{x}_i(t + \Delta t) \in \mathbb{R}$ | The predicted state of node *i* at time *t*+Δ*t* |
| $a_k, b_k, c_\mathrm{k}, d_k, e_k$ | The community-level coefficients for community *k* |
| $\phi_i, \varphi_i$ | The individual-level coefficients for node *i* |

### *1.2 Candidate-function catalog and adaptive library roles in SAGE*

The initial candidate library serves only as a compact active search space and contains a small set of common low-complexity dynamical terms. It does not exhaustively enumerate all candidate terms that may be considered during discovery. Instead, SAGE maintains a broader catalog of admissible function families and mathematical primitives that defines the structural prior for candidate-space refinement. During subsequent repair rounds, the LLM analyzes residual patterns together with review feedback to determine which dynamical mechanisms are insufficiently represented in the current library. Based on this diagnosis, SAGE can activate previously unused function families and dynamically instantiate, reparametrize, or compose additional candidate terms from the admissible primitives, including higher-order nonlinearities, state–time interactions, and topology-aware interaction structures. The resulting candidates are incorporated into the active library only after redundancy and collinearity screening. In this way, the candidate space evolves according to the observed failure modes of the current equation rather than being fully enumerated in advance.

Table S2. Candidate-function families and their roles in the adaptive library of SAGE

| Function family | Symbol | Library role | Description |
|---|---|---|---|
| Constant | $1$ | Base | Baseline term or constant external drive |
| Self-dynamics | $x, y, z$ | Base | Linear self-dependent dynamics of node (i) |
| Polynomial self-interaction | $x_i^2, x_i x_j$ | Base + adaptive extension | Second-order nonlinear state interactions within node states |
| Cubic self-interaction | $x^2, xy, xyz, x^3$ | Adaptive expansion | Higher-order nonlinear self-effects and multi-dimensional interactions |
| Trigonometric | $\sin(x_i), \cos(x_i)$ | Base + adaptive extension | Oscillatory or periodic nonlinear dynamics |
| Exponential / decay | $e^{x_i}$ | Adaptive expansion | Exponential growth, decay, or relaxation mechanisms |
| Network mean coupling | $\boldsymbol{A}x$ | Base | Neighbor averaging and global diffusion-like coupling |
| Network diffusion | $\boldsymbol{A}(x_j - x_i)$ | Base | Difference-based diffusive interactions between connected nodes |
| Nonlinear network coupling | $\boldsymbol{A}f(x_j - x_i)$ | Adaptive expansion | Nonlinear topology-aware interactions through network connections |

| Function family | Symbol | Library role | Description |
|---|---|---|---|
| Network autoregressive coupling | $\boldsymbol{A}f(t-\tau)$ | Adaptive expansion | Delayed network effects and temporal neighbor dependence |
| Saturation | $\tanh(x_i), sigmoid(x_i)$ | Base + adaptive extension | Saturating or threshold-like responses |
| Explicit time dependence | $t, \sin(\omega t), \cos(\omega t)$ | Base + adaptive extension | Time-varying external forcing and periodic driving |
| State-time modulation | $x_i\sin(\omega t), x_i\cos(\omega t)$ | Adaptive expansion | Time modulation of state-dependent dynamics |
| Time-state interaction | $tx_i, \sin(t)x_i$ | Adaptive expansion | Coupled temporal and state-dependent effects |
| Autoregressive terms | $x_i(t-\tau)$ | Adaptive expansion | Short-term memory and delayed state effects |
| Moving-average terms | $M\boldsymbol{A}_\omega(x_i)$ | Adaptive expansion | Smoothed historical dependence over temporal windows |
| …… | …… | Adaptive expansion | …… |

### *1.3 Adaptive candidate function library expansion*

During equation discovery, the initial candidate-function library may not contain all dynamical mechanisms required to describe the observed system. Instead of manually specifying an excessively large function library, SAGE enables adaptive library expansion guided by residual diagnostics from the Review Agent. When the selected equation fails to explain specific dynamical patterns, the Review Agent summarizes the remaining errors and identifies potential missing mechanisms. Based on these diagnostic signals, the LLM identifies appropriate function families from the broader structural catalog and determines whether existing candidate terms should be activated, reparametrized, or extended through admissible compositions of mathematical primitives.

At search round r, let $S(REPAIR^{(r)})$ denote the set of function families selected for repair. Each selected function family is translated into concrete candidate terms through a term-generation operator, which may instantiate predefined forms or construct admissible higher-order, interaction, or parameterized variants from the available primitives.

$$S(REPAIR^{(r)}) = \{L_1, L_2, \dots, L_q\} \quad \text{(S1)}$$

where $S(\cdot)$ generates explicit mathematical candidate terms from the selected function families, such as polynomial, periodic, saturation, or interaction terms, which are subsequently added to the candidate function library. The expanded library is then obtained by combining the existing candidate-function library with the newly generated terms:

$$L^{(r)} \cup S(REPAIR^{(r)}) \quad \text{(S2)}$$

To prevent uncontrolled growth of the candidate library, a pruning operation is applied:

$$L^{(r+1)} = P[L^{(r)} \cup S(REPAIR^{(r)})] \quad \text{(S3)}$$

The pruning procedure removes redundant, duplicated, or highly correlated candidate terms while preserving function families that provide complementary explanatory capabilities. The selection of additional function families is determined according to the characteristics of the residual dynamics. The Review Agent analyzes residual patterns from three complementary perspectives: temporal dynamics, state dependence, and interaction effects. These diagnostic patterns are converted into structured feedback for the LLM, which selects corresponding function families from the predefined catalog. Typical mappings between residual patterns and candidate function families are summarized below:

Table S3. Mapping between residual patterns, candidate function families, and corresponding candidate terms

| Residual pattern | Candidate function families | Candidate function term |
|---|---|---|
| Periodic oscillation in residuals | Oscillatory dynamics | sin(x), cos(x) |
| Nonlinear amplitude variation | Nonlinear self-interaction | Polynomial terms |
| Saturating response | Bounded nonlinear effects | Sigmoid, hyperbolic tangent functions |
| Exponential growth or decay | Growth/attenuation mechanisms | Exponential terms |
| Persistent cross-node dependency | Network interaction mechanisms | Topology-aware interaction terms |
| Time-dependent residual drift | Explicit temporal effects | Time-dependent basis functions |

### *1.4 Matrix representation of SAGE*

To provide a compact mathematical representation of SAGE, we describe its equation-discovery procedure in matrix form. Consider a complex network with *N* nodes, clustered into *K* communities. The task is to infer the governing equation of each node from its observed state trajectories. We assume that the governing dynamics are sparse in the current candidate-function space, such that only a small subset of candidate terms is required to explain the observed dynamics. For simplicity, we consider one state dimension $x_i(t)$ of node *i* in community *k*. The procedure for other dimensions is the same. The dynamics of this state dimension are decomposed into a community-level component and an individual-level component:

$$\begin{aligned}\frac{\mathrm{d}x_i(t)}{\mathrm{d}t} &= S_k(x_i(t)) + H_i(x_i(t)) \\ &= \mathbf{\Theta}^{(r)}(x_i(t))\boldsymbol{\xi}_k^{S(r)} + \mathbf{\Theta}^{(r)}(x_i(t))\boldsymbol{\xi}_i^{H(r)} \\ &= [1, x_i(t), x_i^2(t), \sin(x_i(t)), \cdots]\begin{bmatrix}\xi_{k,1}^{S(r)} \\ \vdots \\ \xi_{k,M}^{S(r)}\end{bmatrix} + [1, x_i(t), x_i^2(t), \sin(x_i(t)), \cdots]\begin{bmatrix}\xi_{i,1}^{H(r)} \\ \vdots \\ \xi_{i,M}^{H(r)}\end{bmatrix}\end{aligned} \quad \text{(S4)}$$

where $\mathbf{\Theta}^{(r)}(x_i(t)) \in \mathbb{R}^{M_r}$ denotes the candidate-function vector available at search round r, and $M_r$ is the number of candidate terms in the current library. The library may contain self-dynamics, nonlinear state interactions, periodic terms and topology-aware interactions, as described in Section 1.2. The coefficient vector $\boldsymbol{\xi}_K^S \in \mathbb{R}^{M_r}$ represents the community-level dynamics shared by nodes in community *k*, whereas $\boldsymbol{\xi}_i^H \in \mathbb{R}^{M_r}$ represents the additional individual-level dynamics specific to node $i$.

**(1) Community-level dynamics**

We first infer the community-level dynamics shared by the $N_k$ nodes in community $k$. For the considered state dimension, the observed trajectories of these nodes are arranged into the state matrix

$$\boldsymbol{X}_k = \begin{bmatrix} x_1(1) & x_1(2) & \cdots & x_1(T) \\ x_2(1) & x_2(2) & \cdots & x_2(T) \\ \vdots & \vdots & \vdots & \vdots \\ x_{N_k}(1) & x_{N_k}(2) & \cdots & x_{N_k}(T) \end{bmatrix} \in \mathbb{R}^{N_k \times T} \rightarrow \boldsymbol{x}_k = \begin{bmatrix} x_1(1) \\ \vdots \\ x_1(T) \\ x_2(1) \\ \vdots \\ x_2(T) \\ \vdots \\ x_{N_k}(T) \end{bmatrix} \in \mathbb{R}^{N_k T} \quad \text{(S5)}$$

The corresponding temporal derivatives are arranged in the same order:

$$\frac{\mathrm{d}\boldsymbol{x}}{\mathrm{d}t} = \begin{bmatrix} \frac{\mathrm{d}x_1(1)}{\mathrm{d}t} & \frac{\mathrm{d}x_1(2)}{\mathrm{d}t} & \cdots & \frac{\mathrm{d}x_1(T)}{\mathrm{d}t} \\ \frac{\mathrm{d}x_2(1)}{\mathrm{d}t} & \frac{\mathrm{d}x_2(2)}{\mathrm{d}t} & \cdots & \frac{\mathrm{d}x_2(T)}{\mathrm{d}t} \\ \vdots & \vdots & \vdots & \vdots \\ \frac{\mathrm{d}x_{N_k}(1)}{\mathrm{d}t} & \frac{\mathrm{d}x_{N_k}(2)}{\mathrm{d}t} & \cdots & \frac{\mathrm{d}x_{N_k}(T)}{\mathrm{d}t} \end{bmatrix} \in \mathbb{R}^{N_k \times T} \rightarrow \frac{\mathrm{d}\boldsymbol{x}}{\mathrm{d}t} = \begin{bmatrix} \frac{\mathrm{d}x_1(1)}{\mathrm{d}t} \\ \vdots \\ \frac{\mathrm{d}x_1(T)}{\mathrm{d}t} \\ \frac{\mathrm{d}x_2(1)}{\mathrm{d}t} \\ \vdots \\ \frac{\mathrm{d}x_2(T)}{\mathrm{d}t} \\ \vdots \\ \frac{\mathrm{d}x_{N_k}(T)}{\mathrm{d}t} \end{bmatrix} \in \mathbb{R}^{N_k T} \quad \text{(S6)}$$

The candidate functions in the current library $L^{(r)}$ are evaluated at each node–time sample to construct the community-level design matrix $\Theta_k^{(r)}$.

$$\boldsymbol{\Theta}_k^{(r)}(x) = \begin{bmatrix} 1 & x_1(1) & x_1^2(1) & \sin(x_1(1)) & \cdots \\ \vdots & \vdots & \vdots & \vdots & \vdots \\ 1 & x_1(T) & x_1^2(T) & \sin(x_1(T)) & \cdots \\ 1 & x_2(1) & x_2^2(1) & \sin(x_2(1)) & \cdots \\ \vdots & \vdots & \vdots & \vdots & \vdots \\ 1 & x_2(T) & x_2^2(T) & \sin(x_2(T)) & \cdots \\ \vdots & \vdots & \vdots & \vdots & \vdots \\ 1 & x_{N_k}(T) & x_{N_k}^2(T) & \sin\left(x_{N_k}(T)\right) & \cdots \end{bmatrix} \in \mathbb{R}^{N_k T \times M_r} \tag{S7}$$

For a candidate community-level coefficient vector $\xi_k^S$ , the stacked dynamics are represented as

$$\begin{bmatrix} \frac{\mathrm{d}x_1(1)}{\mathrm{d}t} \\ \vdots \\ \frac{\mathrm{d}x_1(T)}{\mathrm{d}t} \\ \frac{\mathrm{d}x_2(1)}{\mathrm{d}t} \\ \vdots \\ \frac{\mathrm{d}x_2(T)}{\mathrm{d}t} \\ \vdots \\ \frac{\mathrm{d}x_{N_k}(T)}{\mathrm{d}t} \end{bmatrix} = \begin{bmatrix} 1 & x_1(1) & x_1^2(1) & \sin(x_1(1)) & \cdots \\ \vdots & \vdots & \vdots & \vdots & \vdots \\ 1 & x_1(T) & x_1^2(T) & \sin(x_1(T)) & \cdots \\ 1 & x_2(1) & x_2^2(1) & \sin(x_2(1)) & \cdots \\ \vdots & \vdots & \vdots & \vdots & \vdots \\ 1 & x_2(T) & x_2^2(T) & \sin(x_2(T)) & \cdots \\ \vdots & \vdots & \vdots & \vdots & \vdots \\ 1 & x_{N_k}(T) & x_{N_k}^2(T) & \sin\left(x_{N_k}(T)\right) & \cdots \end{bmatrix} \begin{bmatrix} \xi_{k,1}^{S(r)} \\ \xi_{k,2}^{S(r)} \\ \vdots \\ \xi_{k,M}^{S(r)} \end{bmatrix} \tag{S8}$$

The objective function of this sparse regression problem is given as:

$$\boldsymbol{E}_K(\xi) = \frac{1}{N_K T} \left\| x_K - \Theta_k^{(r)} \xi \right\|_2^2, \qquad C(\xi) = \|\xi\|_0 \tag{S9}$$

Bootstrap STLSQ, cross-validated Lasso and Ridge regression under multiple threshold settings generate multiple candidate coefficient vectors and support sets from Eq. (S9). These candidates are compared using validation error $E_K$ , structural complexity $C$ and stability evidence derived from cross-solver consistency, bootstrap resampling, and coefficient sign consistency. Pareto filtering retains candidate structures that are not simultaneously dominated in validation accuracy and structural complexity, after which the selected community-level structure is refitted numerically to obtain $\xi_k^S$.The non-zero entries of $\xi_k^S$ define the community-level terms shared by nodes in community $K$.

**(2) Individual-level dynamics**

After identifying the community-level structure, SAGE examines the residual dynamics of each node to determine whether additional individual-level terms are required. For node $i$ in community $K$, the trajectory of the considered state dimension is written as:

$$\boldsymbol{x}_i(t) = [x_i(1) \quad x_i(2) \quad \cdots \quad x_i(T)]^T \in \mathbb{R}^T \tag{S10}$$

Its temporal derivatives are arranged as:

$$\frac{\mathrm{d}\boldsymbol{x}_{\mathrm{i}}(t)}{\mathrm{d}t}=\begin{bmatrix}\frac{\mathrm{d}x_i(1)}{\mathrm{d}t} & \frac{\mathrm{d}x_i(2)}{\mathrm{d}t} & \cdots & \frac{\mathrm{d}x_i(T)}{\mathrm{d}t}\end{bmatrix}^T \in \mathbb{R}^T \tag{S11}$$

The current candidate library is evaluated along this trajectory to form the node-specific design matrix:

$$\boldsymbol{\Theta}_k^{(r)}(x)=\begin{bmatrix}1 & x_i(1) & x_i^2(1) & \sin(x_i(1)) & \cdots\\ 1 & x_i(2) & x_i^2(2) & \sin(x_i(2)) & \cdots\\ \vdots & \vdots & \vdots & \vdots & \vdots\\ 1 & x_i(T) & x_i^2(T) & \sin(x_i(T)) & \cdots\end{bmatrix} \in \mathbb{R}^{T\times M_r} \tag{S12}$$

After subtracting the fitted community-level contribution, the unexplained dynamics of node $i$ are:

$$Rse_i=\boldsymbol{x}_i-\boldsymbol{\Theta}_i^{(r)}\boldsymbol{\xi}_k^{S(r)}\ ,\ Rse_i\in\mathbb{R}^T$$

$$\begin{bmatrix}Rse_i(1)\\ Rse_i(2)\\ \vdots\\ Rse_i(T)\end{bmatrix}=\begin{bmatrix}\frac{\mathrm{d}x_i(1)}{\mathrm{d}t}\\ \frac{\mathrm{d}x_i(2)}{\mathrm{d}t}\\ \vdots\\ \frac{\mathrm{d}x_i(T)}{\mathrm{d}t}\end{bmatrix}-\begin{bmatrix}1 & x_i(1) & x_i^2(1) & \sin(x_i(1)) & \cdots\\ 1 & x_i(2) & x_i^2(2) & \sin(x_i(2)) & \cdots\\ \vdots & \vdots & \vdots & \vdots & \vdots\\ 1 & x_i(T) & x_i^2(T) & \sin(x_i(T)) & \cdots\end{bmatrix}\begin{bmatrix}\xi_{k,1}^{S(r)}\\ \xi_{k,2}^{S(r)}\\ \vdots\\ \xi_{k,M}^{S(r)}\end{bmatrix} \tag{S13}$$

Individual-level terms are then used to explain this residual:

$$\boldsymbol{r}_i=\boldsymbol{\Theta}_i^{(r)}\boldsymbol{\xi}_i^{H(r)} \tag{S14}$$

A non-zero entry in $\xi_i^H$ may therefore represent either an additional node-specific function or a node-specific coefficient adjustment to a function already present in the community-level structure.

Combining the selected community-level and individual-level components gives the governing equation of node $i$ in community $k$ :

$$\frac{\mathrm{d}x_i(t)}{\mathrm{d}t}=S_k\big(x_i(t)\big)+H_i\big(x_i(t)\big)=\boldsymbol{\Theta}^{(r)}\big(x_i(t)\big)\hat{\boldsymbol{\xi}}_k^S+\boldsymbol{\Theta}^{(r)}\big(x_i(t)\big)\hat{\boldsymbol{\xi}}_i^H \tag{S15}$$

### *1.5 Pseudocode of SAGE*

Algorithm 1 summarizes the complete Library–Discovery–Review procedure of SAGE for governing-equation discovery. The framework first initializes the candidate function library from the observed node-state observations and then iteratively searches, evaluates, and refines governing equations. The Library Agent constructs and updates the candidate function library based on residual patterns and diagnostic feedback. The Discovery Agent identifies sparse equation structures from the current library using numerical regression and selects candidate structures according to validation error, structural complexity, and stability evidence. The Review Agent evaluates the selected equation using derivative fit, coefficient stability, and state prediction, and provides repair instructions when the equation fails the predefined criteria.

During the iterative search, all adaptive operations, including candidate-library updates, equation selection, and repair decisions, are conducted using only the training and validation sets. The test set is completely isolated from the search process and is used only for the final evaluation after an equation receives

ACCEPT. All numerical configurations, search parameters, and random seeds are specified in a unified configuration file to ensure reproducibility across experiments.

**Algorithm 1** | LLM-SAGE for governing-equation discovery in community-structured complex networks

**Input:**
$\{\mathbf{x}_i(t) \mid i = 1, \ldots, N;\ t = 1, \ldots, T\}$: node-state observations
$\mathbf{A}$: adjacency matrix, when available
$\mathbf{C}$: community labels, provided or inferred from node-state trajectories
$R_{\max}$: maximum number of repair rounds
**Output:** $\mathbf{S}_k(\mathbf{x}_i(t), \mathbf{A})$, $H_i(\mathbf{x}_i(t), \mathbf{A})$, and final governing equations
*# Data preprocessing and initialization*
Preprocess observations, estimate temporal derivatives, and split data into training, validation, and test sets.
Determine community assignments and construct initial candidate library $\mathcal{L}^{(0)}$.
**for** $r = 0$ to $R_{\max}$ **do**
    Construct candidate-function vectors $\Theta^{(r)}(\mathbf{x}_i(t))$ from $\mathcal{L}^{(r)}$ and assemble community-level design matrices $\Theta_k^{(r)}$.
    The Discovery Agent generates sparse candidate equations using bootstrap STLSQ, cross-validated Lasso, and validation-guided Ridge sparsification.
    Aggregate stability evidence and construct Pareto candidate equations.
    Select community-level support structures and refit coefficients $\xi_k^{S(r)}$.
    Compute residual dynamics

$$\mathrm{Res}_i^{H(r)}(t) = \frac{d\mathbf{x}_i(t)}{dt} - \Theta^{(r)}(\mathbf{x}_i(t))\, \xi_k^{S(r)}$$

    and identify individual-level candidate terms.
    Combine community-level and individual-level support sets, re-estimate $\xi_k^{S(r)}$ and $\xi_i^{H(r)}$, and simplify the model.
    The Review Agent evaluates derivative fit, coefficient stability, and one-step state prediction.
    **if** ACCEPT **then**
        Evaluate on independent test set and return final equations.
    **else if** REPAIR **and** $r < R_{\max}$ **then**
        Map repair plan to function families $\mathcal{R}^{(r)}$ and update:

$$\mathcal{L}^{(r+1)} = \mathcal{P}\left[\mathcal{L}^{(r)} \cup \mathcal{S}\left(\mathcal{R}^{(r)}\right)\right]$$

    **else**
        Return REJECT and diagnostic report.
    **end if**
**end for**
*# Note*
All candidate-library updates, model-selection steps, and repair decisions use only the training and validation sets. The test set is excluded from all adaptive decisions until a model is accepted.

### *1.6 LLM prompts and structured returns*

To illustrate how SAGE integrates language-mediated decisions into the numerical equation-discovery pipeline, we provide a representative execution trace on a real-world transportation network. This example records the structured information provided to each LLM agent, the corresponding prompts, and the returned decisions during one complete Library–Discovery–Review repair cycle. The following example demonstrates how language-based decisions interact with numerical equation discovery.

**Library Agent: residual diagnosis and function-family expansion**

This call corresponds to the adaptive library update described in Library Agent. The Review Agent provides diagnostic feedback on the current equation, including unexplained residual patterns and unstable terms. Based on this feedback, the Library Agent selects additional or revised function families from the predefined catalog to update the candidate library.

Table S4. LLM interaction trace of the Library Agent

| Representative prompt | Example structured return |
| --- | --- |
| Representative prompt<br>Agent role: You are the Library Agent in SAGE.<br>Input information provided by the numerical module:<br>Residual diagnostics:<br>residual_std_ratio: relative residual magnitude<br>lag1_autocorr: temporal correlation of residuals<br>dominant_frequency: dominant frequency components detected from residual spectra<br>base_term_correlations: correlations between residuals and current candidate terms<br>network_residual_correlation: correlation between residuals and network topology<br>community_residual_stats: residual statistics across communities<br>Current candidate-function library: Existing function families and candidate terms used in the current search round.<br>Allowed expansion families: Function families that can be added from the predefined candidate function catalog.<br>Task:<br>Diagnose whether the current candidate library lacks important dynamical mechanisms.<br>Identify the most plausible missing function families based only on the provided diagnostics.<br>Select function families only from the allowed expansion catalog.<br>Do not generate governing equations, mathematical expressions, or numerical coefficients.<br>Return a structured JSON response following the required format.<br>Required JSON output format: | {<br>"diagnosis": [<br>"persistent oscillatory residual patterns remain after the current library fitting"<br>],<br>"recommended_families": [<br>"trigonometric",<br>"state_time_modulation"<br>],<br>"evidence": [<br>"dominant_frequency",<br>"high_lag1_autocorr",<br>"residual_pattern_not_explained_by_current_terms"<br>],<br>"rationale": "The residual dynamics exhibit stable periodic behavior that is not captured by the current candidate library. Adding trigonometric and state-time modulation terms may provide additional mechanisms to represent the observed temporal variation."<br>} |

| Representative prompt | Example structured return |
|---|---|
| {<br>"diagnosis": [ "brief description of unexplained dynamical patterns"],<br>"recommended_families": [ "selected function families from the allowed catalog"],<br>"evidence": [ "diagnostic signals supporting the decision"],<br>"rationale": "brief explanation of why these families may explain the residual dynamics"<br>} | |

**Discovery Agent: Pareto candidate structure selection**

This call corresponds to the Pareto candidate selection described in Discovery Agent. The numerical module first generates candidate equations using Bootstrap STLSQ, Lasso, and Ridge regression. These candidates are evaluated using validation error, structural complexity, and stability evidence. The Discovery Agent then compares the Pareto candidates and selects the most appropriate equation structure.

Table S5. LLM interaction trace of the Discovery Agent

| Representative prompt | Example structured return |
|---|---|
| Representative prompt<br>Agent role: You are the Discovery Agent in SAGE.<br>Input information provided by the numerical module:<br>Pareto candidate equations: Candidate structures generated by sparse regression methods and retained after Pareto filtering.<br>Candidate evaluation evidence:<br>validation_error: prediction error on the validation set<br>complexity: number of retained candidate terms<br>stability_matrix: term selection consistency across numerical solvers<br>bootstrap_stability: selection frequency from bootstrap resampling<br>sign_consistency: coefficient sign consistency across repeated estimations<br>Term metadata: Function-family information and physical interpretation labels of candidate terms.<br>Task:<br>Compare only the supplied Pareto candidate structures.<br>Select the candidate equation that provides the best trade-off between accuracy, sparsity, and stability. | {<br>"decision": "SELECT",<br>"model_id": "Pareto-03",<br>"supported_terms": [<br>"self_linear",<br>"polynomial_self",<br>"network_diffusion"<br>],<br>"concerns": [<br>"weak evidence for additional nonlinear coupling terms"<br>],<br>"rationale": "Pareto-03 achieves a favorable balance between validation accuracy, model sparsity, and cross-solver stability. The retained terms are consistently supported by Bootstrap STLSQ, LassoCV, and Ridge threshold scanning."<br>} |

| Representative prompt | Example structured return |
| --- | --- |
| Prefer structures supported by multiple numerical solvers and repeated estimations.<br>Do not add new candidate terms.<br>Do not modify or estimate numerical coefficients.<br>Return a structured JSON response following the required format.<br>Required JSON output format:<br>{<br>"decision": "SELECT\|REVISE",<br>"model_id": "identifier of selected Pareto candidate",<br>"supported_terms": [<br>"retained candidate terms"<br>],<br>"concerns": [<br>"remaining limitations or uncertainties"<br>],<br>"rationale": "brief explanation of the selection decision"<br>} | } |

**Implementation note:** The LLM only performs structure-level adjudication among existing Pareto candidates, whereas coefficient estimation and model reconstruction are performed by numerical regression.

**Review Agent: reflective review and repair**

This call corresponds to the review stage described in Review Agent. The Review Agent evaluates the selected equation using derivative fit, coefficient stability, and state prediction. Based on these criteria, it returns ACCEPT, REPAIR, or REJECT. When REPAIR is returned, the Review Agent provides diagnostic information that guides the next library update round.

Table S6. LLM interaction trace of the Review Agent

| Representative prompt | Example structured return |
| --- | --- |
| Agent role: You are the Review Agent in SAGE.<br>Input information provided by the numerical module:<br>Current equation structure: Selected sparse governing equation and retained function terms.<br>Evaluation evidence:<br>derivative_fit: agreement between observed and predicted derivatives | {<br>"decision": "REPAIR",<br>"failed_checks": [<br>"state_prediction",<br>"structured_residual"<br>], |

| Representative prompt | Example structured return |
|---|---|
| bootstrap_stability: coefficient uncertainty estimated from bootstrap resampling<br>state_prediction: step prediction performance on the validation set<br>residual_diagnostics: remaining residual patterns and unexplained structures<br>Task:<br>Evaluate whether the current equation satisfies the predefined review criteria.<br>Return exactly one decision: ACCEPT, REPAIR, or REJECT.<br>If ACCEPT, indicate that the equation can proceed to independent test evaluation.<br>If REPAIR, identify the failed evaluation aspects and recommend function families that may capture the missing dynamics.<br>If REJECT, summarize the main reasons why a stable equation cannot be supported.<br>Do not modify numerical coefficients or generate new governing equations.<br>Return a structured JSON response following the required format.<br>Required JSON output format:<br>{<br>"decision": "ACCEPT\|REPAIR\|REJECT",<br>"failed_checks": [<br>"failed evaluation criteria"<br>],<br>"failure_mode": "description of remaining problem",<br>"repair_families": [<br>"recommended function families when REPAIR is selected"<br>],<br>"rationale": "brief explanation of the decision"<br>} | "failure_mode": "The equation reproduces local derivative patterns but leaves a persistent oscillatory component that affects future state prediction.",<br>"repair_families": [<br>"trigonometric",<br>"state_time_modulation"<br>],<br>"rationale": "Residual diagnostics indicate a stable periodic pattern not explained by the current candidate library. Additional oscillatory function families are recommended for the next search round."<br>} |

When ACCEPT is returned, the selected equation is transferred to the independent test set for final evaluation. When REPAIR is returned, the identified failure patterns and recommended function families are passed to the Library Agent, which expands or modifies the candidate function library according to Equation 12. When REJECT is returned, the search terminates and a diagnostic report is generated to summarize the remaining limitations, failed evaluations, and iteration history.

**LLM configuration.**

All language-model calls were made through the gpt-5.6-sol API using the chat-completions interface in JSON mode. The temperature was set to 0.2 to reduce response variability, and the request timeout was 60 s. The maximum number of repair rounds was set to five. All LLM outputs were required to follow the corresponding JSON schema before entering the numerical pipeline.

# 2. Supplementary details and results for synthetic complex network

### *2.1 Supplementary details for the single-dynamics network*

The Lotka–Volterra system (predator–prey model), originally proposed by Alfred J. Lotka and Vito Volterra, is a classical nonlinear dynamical system widely used to describe interactions among two or more populations in ecological systems. It is particularly suited for modeling ecological processes such as predation, competition, and symbiosis. In this study, we extend the Lotka–Volterra dynamics to a community-structured complex network. The network consists of 60 nodes, where each node represents an ecological unit and its state represents the corresponding population size. Network topology is generated using the Stochastic Block Model (SBM) and clustered into three communities. Nodes within the same community have a higher connection probability, whereas connections between different communities are less frequent. The simulation runs for a total duration of 30 with a time step of 0.01. The governing equations are given as follows:

$$\begin{cases} \frac{\mathrm{dx}_i}{\mathrm{d}t} = \alpha_k x_\mathrm{i} - \beta_k x_i y_i + \varepsilon_\mathrm{k} \sum_{j \in \mathbb{N}_i} x_j + \zeta_i \sin(1.5t) x_i \\ \frac{\mathrm{d}y_i}{\mathrm{d}t} = \delta_k x_i y_i - \gamma_k y_i + \varsigma_i y_i \end{cases} \tag{S16}$$

where $\alpha_k, \beta_k, \varepsilon_\mathrm{k}, \delta_k, \gamma_k$ are community-level coefficients for node *i* within community *k*. All nodes within community k share the same community-level coefficients, which characterize their consistent dynamic behavior. In contrast, different communities are assigned to distinct and significantly different coefficients, highlighting dynamical differences across communities. $\zeta_i, \varsigma_i$ are individual-level coefficients for node *i* and represent variation at the individual level. Each node carries distinct coefficients that highlight its unique dynamic behavior. $\varepsilon_\mathrm{k} \sum_{j \in \mathbb{N}_i} x_j$ denotes the interaction effect of the neighboring nodes on node $i$. The detailed dynamics of the three communities in the single-dynamics network are as follows.

$$
\begin{cases}
\text{Group1:}\begin{cases}
\dfrac{\mathrm{dx}_i}{\mathrm{d}t} = 2x_i - 1x_i y_i + 0.3\sum_{j\in\mathbb{N}_i} x_j + \zeta_i x_i \\
\dfrac{\mathrm{d}y_i}{\mathrm{d}t} = 0.5x_i y_i - y_i + \varsigma_i y_i
\end{cases} \\
\text{Group2:}\begin{cases}
\dfrac{\mathrm{dx}_i}{\mathrm{d}t} = 1.5x_i - 0.8x_i y_i + 0.3\sum_{j\in\mathbb{N}_i} x_j + \zeta_i x_i \\
\dfrac{\mathrm{d}y_i}{\mathrm{d}t} = 0.4x_i y_i - 0.8y_i + \varsigma_i y_i
\end{cases} \\
\text{Group3:}\begin{cases}
\dfrac{\mathrm{dx}_i}{\mathrm{d}t} = 3x_i - 1.5x_i y_i + 0.3\sum_{j\in\mathbb{N}_i} x_j + \zeta_i x_i \\
\dfrac{\mathrm{d}y_i}{\mathrm{d}t} = 0.6x_i y_i - 1.2y_i + \varsigma_i y_i
\end{cases}
\end{cases}
\tag{S17}
$$

where $\xi_i$ and $\varsigma_i$ represent individual-level coefficients of each node, which follow a normal distribution with mean 0 and variance 0.1. Figure S1 illustrates the inference results across the three communities.

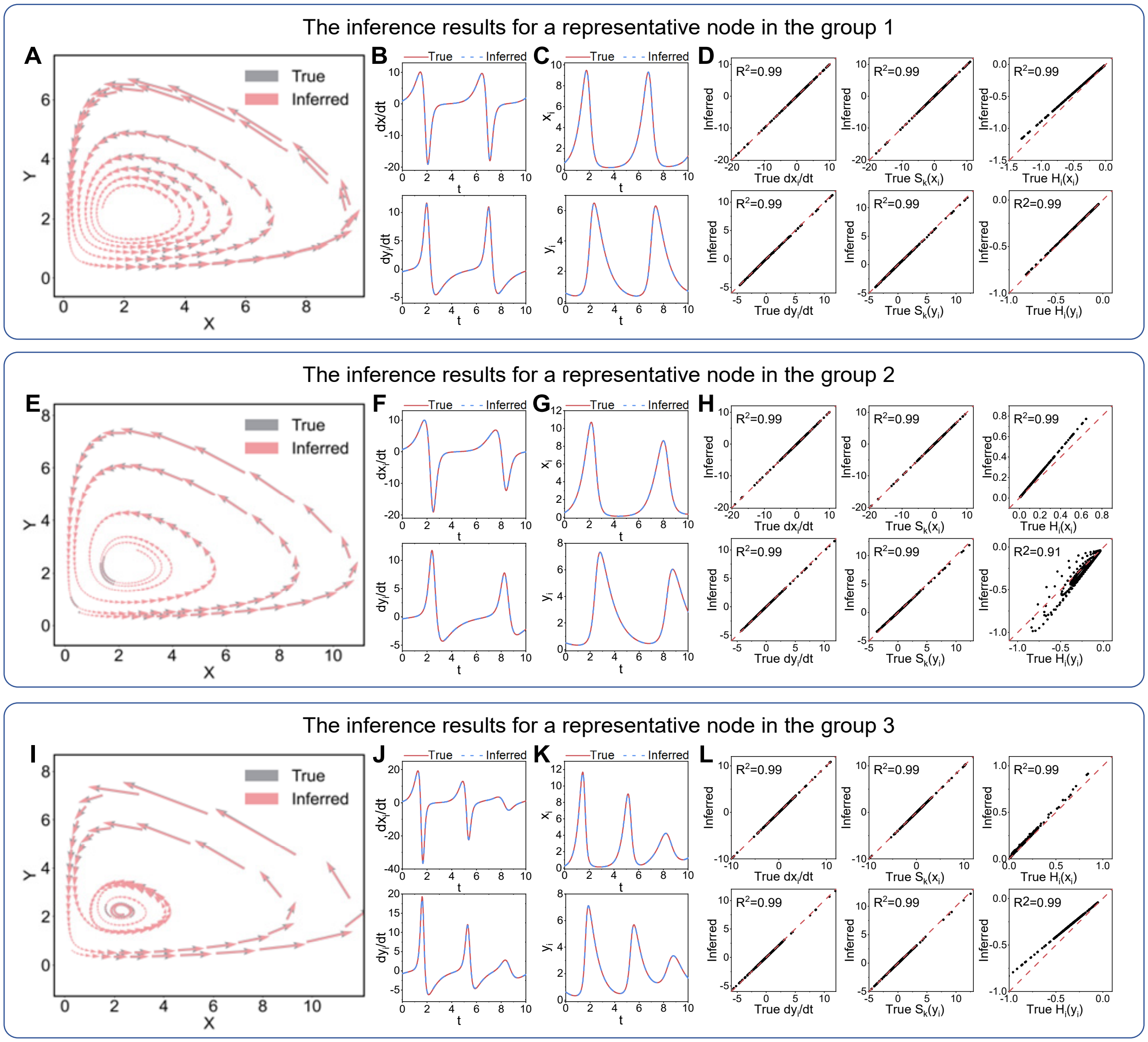

**Figure S1. Validating SAGE on a single-dynamic network with 60 nodes divided into three communities.** A-L Inference results for three representative nodes, one randomly selected from each community.

### *2.2 Supplementary details for the multi-dynamics network*

Chaotic systems, as typical nonlinear dynamical systems, are characterized by strong nonlinearity, sensitivity to initial conditions, non-periodicity, and boundedness. In this study, we extend classical chaotic systems to complex networks and construct a chaotic complex network composed of nodes and edges. The network consists of 60 nodes, with its topological structure generated using the Stochastic Block Model (SBM) and is further clustered into three communities. The nodes in the three communities follow three representative chaotic systems: the Lorenz system, the Rössler system, and the Chen system. The simulation runs for a total duration of 100 with a time step of 0.01. The governing equations are given as follows:

$$
\begin{aligned}
&\text{Lorenz} \begin{cases} \dfrac{\mathrm{d}x_i}{\mathrm{d}t} = a_1(y_i - x_i) + \varsigma_i x_i \\ \dfrac{\mathrm{d}y_i}{\mathrm{d}t} = x_i(b_1 - z_i) - y_i + \zeta_i y_i \\ \dfrac{\mathrm{d}z_i}{\mathrm{d}t} = x_i y_i - c_1 z_i + \xi_i z_i \end{cases} \\
&\text{Rössler} \begin{cases} \dfrac{\mathrm{d}x_i}{\mathrm{d}t} = -y_i - z_i + \varsigma_i z_i \\ \dfrac{\mathrm{d}y_i}{\mathrm{d}t} = x_i + a_2 y_i + \zeta_i y_i \\ \dfrac{\mathrm{d}z_i}{\mathrm{d}t} = b_2 + z_i(x_i - c_2) + \xi_i \end{cases} \\
&\text{Chen} \begin{cases} \dfrac{\mathrm{d}x_i}{\mathrm{d}t} = a_3(y_i - x_i) + \varsigma_i x_i \\ \dfrac{\mathrm{d}y_i}{\mathrm{d}t} = -x_i z_i + c_3 y_i + \zeta_i y_i \\ \dfrac{\mathrm{d}z_i}{\mathrm{d}t} = x_i y_i - b_3 z_i + \xi_i z_i \end{cases}
\end{aligned} \tag{S18}
$$

where $a, b, \mathrm{c}$ are community-level coefficients. Each node follows a governing equation derived from one of the three systems-Lorenz, Rössler, or Chen. $\varsigma_i, \zeta_i, \xi$ are individual-level coefficients for node $i$. The detailed dynamics of the three communities in the multi-dynamics network are given as follows.

$$
\begin{aligned}
&\text{Lorenz} \begin{cases} \dfrac{dx_i}{dt} = 10(y_i - x_i) + \varsigma_i x_i \\ \dfrac{dy_i}{dt} = x_i(28 - z_i) - y_i + \zeta_i y_i \\ \dfrac{dz_i}{dt} = x_i y_i - \dfrac{8}{3} z_i + \xi_i z_i \end{cases} \\
&\text{Rössler} \begin{cases} \dfrac{dx_i}{dt} = -y_i - z_i + \varsigma_i z_i \\ \dfrac{dy_i}{dt} = x_i + 0.3 y_i + \zeta_i y_i \\ \dfrac{dz_i}{dt} = 1 + z_i(x_i - 5.7) + \xi_i \end{cases} \\
&\text{Chen} \begin{cases} \dfrac{dx_i}{dt} = 12(y_i - x_i) + \varsigma_i x_i \\ \dfrac{dy_i}{dt} = -x_i z_i + 8 y_i + \zeta_i y_i \\ \dfrac{dz_i}{dt} = x_i y_i - z_i + \xi_i z_i \end{cases}
\end{aligned} \tag{S19}
$$

Where $\varsigma_i, \zeta_i, \xi_i$ represent the individual-level terms of each node, which follow a normal distribution with mean 0 and variance 0.1. Figure S2 illustrates the inference results across the three communities.

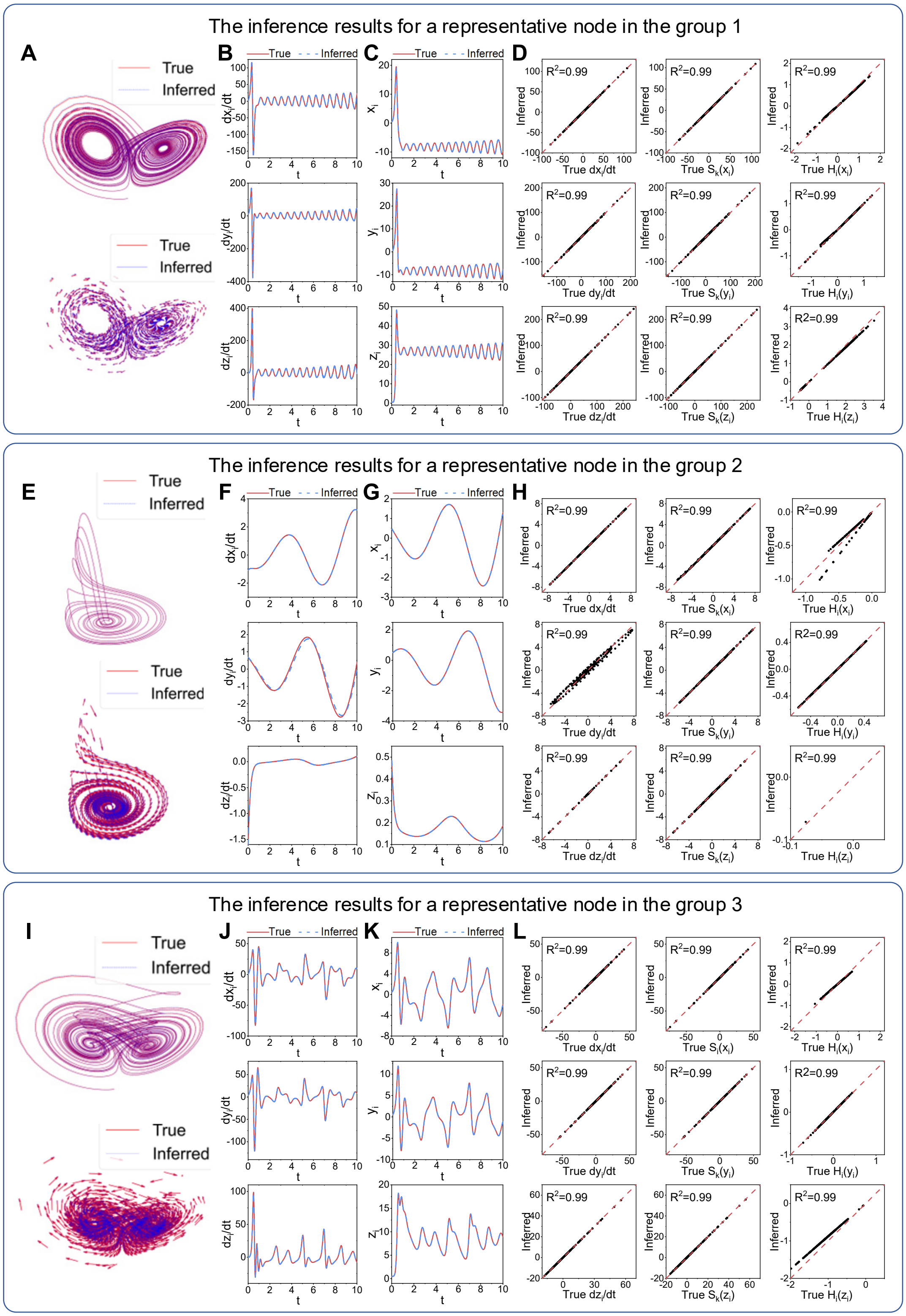

**Figure S2. Validating SAGE on a multi-dynamics network with 60 nodes divided into three communities.** A-L Inference results for three representative nodes, one randomly selected from each community.

## 3. Supplementary Details for Real-World Complex Network Analysis

### *3.1 Urban mobile communication network*

We use the urban mobile communication dataset from the Telecom Italia Big Data Challenge, which records mobile communication activity in Milan, Italy, from November 1 to November 7, 2013, including Internet activity, phone-call activity, and SMS activity. We select 48 spatial grid cells as network nodes. For each node $i$, the state vector is defined as $\boldsymbol{x}_i(t) = [x_i(t), y_i(t), z_i(t)]^{\mathrm{T}}$, where $x_i(t)$,$y_i(t)$, and $z_i(t)$ denote Internet activity, phone-call activity, and SMS activity, respectively. The processed dataset contains 168 steps for each node.

Network topology is constructed according to geographical adjacency. Specifically, two grid cells are connected if they are spatially adjacent under a four-neighbor structure, so that the adjacency matrix represents local interactions among neighboring urban regions. The resulting node-state trajectories and spatial topology are used jointly for hierarchical governing-equation discovery. Detailed inference and interpretation results are presented in the following section.

### *3.2 Brain EEG network*

In this study, we use the EEG Motor Movement Imagery Dataset (EEGMMIDB) from PhysioNet and select the S001R01 recording. The dataset contains 64 EEG channels, which are treated as network nodes. The signals are band-pass filtered between 1 and 20 Hz, re-referenced using common average referencing (CAR), and transformed using the Hilbert transform. For each EEG channel, the analytic signal obtained from the Hilbert transformation is represented by two state variables, where $x_i(t)$ represents the filtered EEG signal, corresponding to the real component of the analytic signal, and $y_i(t) = H[x_i(t)]$ represents the corresponding Hilbert quadrature component. The processed data contains 2,440-time steps and two state variables for each node.

The EEG network topology is constructed from amplitude-envelope functional connectivity between channels. Specifically, the amplitude envelope of each channel is extracted from its analytic signal, and the absolute Pearson correlation coefficient is calculated for every pair of channel envelopes over the entire recording. Among all 2,016 channel pairs, the top 20% with the largest absolute correlations are retained, yielding 403 undirected edges with an empirical cutoff of approximately $|r|$=0.669. The resulting adjacency matrix $\boldsymbol{A}$ is symmetric and binary without self-loops, where $\boldsymbol{A}_{ij} = 1$ indicates strong amplitude-envelope co-fluctuation between channels $i$ and $j$, and $\boldsymbol{A}_{ij} = 0$ otherwise. This topology represents sensor-level functional connectivity rather than anatomical connectivity or spatial-neighborhood relationships.

The 64 EEG channels are organized into five dynamical communities for subsequent equation discovery and hierarchical analysis. These communities are used to identify shared dynamical structures among channels and distinguish community-level mechanisms from channel-specific variations. Detailed governing-equation inference and interpretation results are presented below.

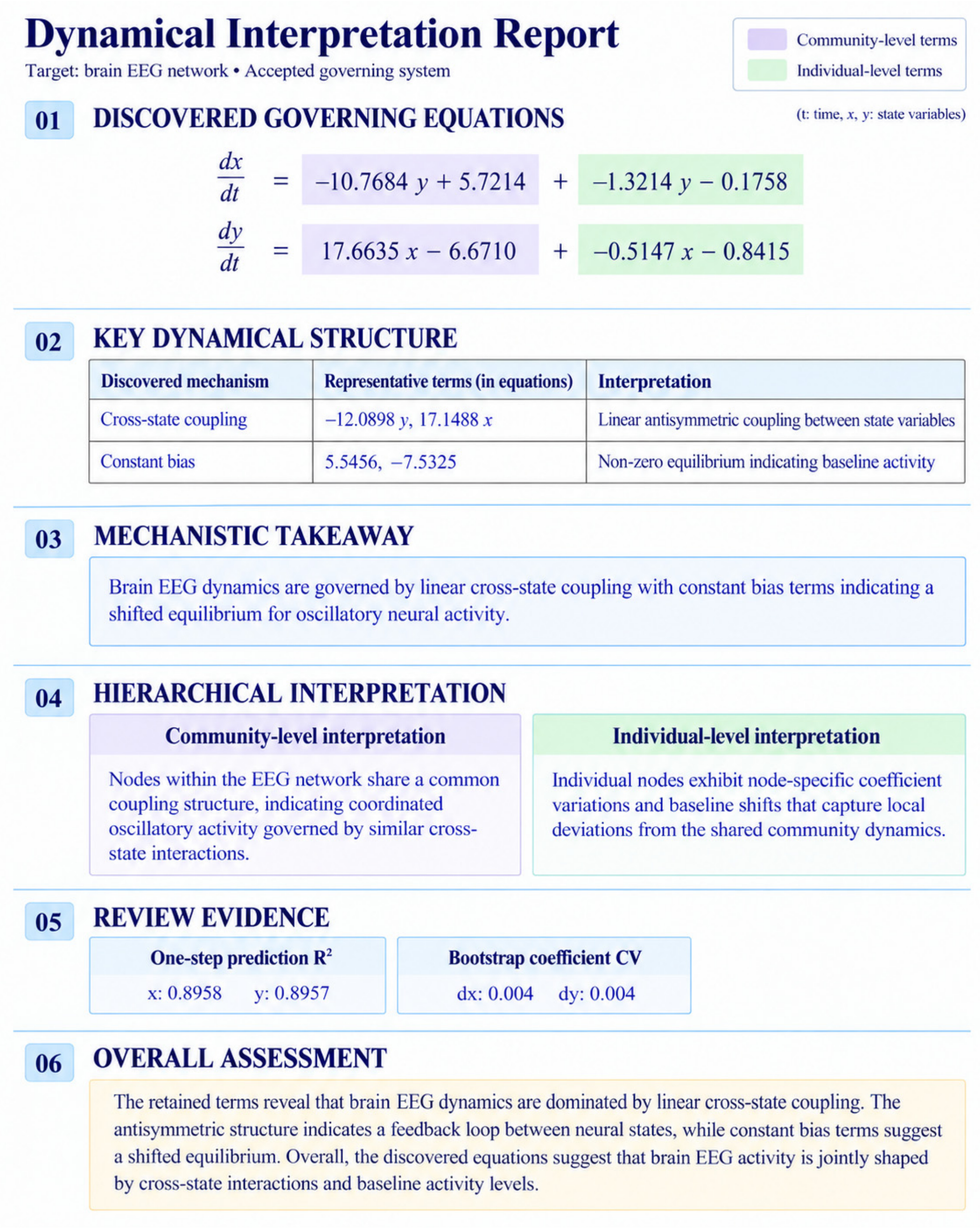

**Figure S3. Dynamical interpretation of the governing equations discovered for the brain EEG network.** The discovered EEG dynamics are dominated by linear cross-state coupling together with constant bias terms. The coupling coefficients reveal an antisymmetric interaction between the two state variables, consistent with feedback-like oscillatory structure, whereas the constant terms indicate a shifted dynamical baseline. The hierarchical decomposition further distinguishes shared community-level coupling patterns from individual-level coefficient variations and baseline shifts. Model reliability is supported by high one-step prediction accuracy and low bootstrap coefficient variation. Purple and green shading denote community-level and individual-level terms, respectively.

### *3.3 Power grid network*

We use the IEEE 39-bus New England power system as a benchmark power-grid network. The system contains 39 buses, which are treated as network nodes and organized into three dynamical communities for hierarchical governing-equation discovery. Each bus $i$ is described by a single state variable $x_i(t)$, representing its phase-angle state associated with rotor-angle dynamics.

The network topology is constructed directly from the physical transmission-line connectivity of the system. Specifically, two buses $i$ and $j$ are considered connected if they are directly linked by a transmission line. The resulting adjacency matrix $\boldsymbol{A}_{ij}$ therefore represents the physical coupling topology among interconnected buses, where $\boldsymbol{A}_{ij} = 1$ indicates a direct transmission-line connection between buses $i$ and $j$, and $\boldsymbol{A}_{ij} = 0$ otherwise. This topology is jointly used with the bus-state trajectories to identify topology-dependent phase-coupling mechanisms in the power-grid dynamics.

The phase-angle trajectories $x_i(t)$ are used as dynamical observations for governing-equation discovery, while the transmission-line adjacency matrix provides the corresponding network-coupling structure. SAGE jointly uses these temporal trajectories and the physical network topology to identify shared and node-specific phase-coupling mechanisms across the power-grid network. Detailed governing-equation inference and dynamical interpretation results are presented below. Detailed governing-equation inference and dynamical interpretation results are presented below.

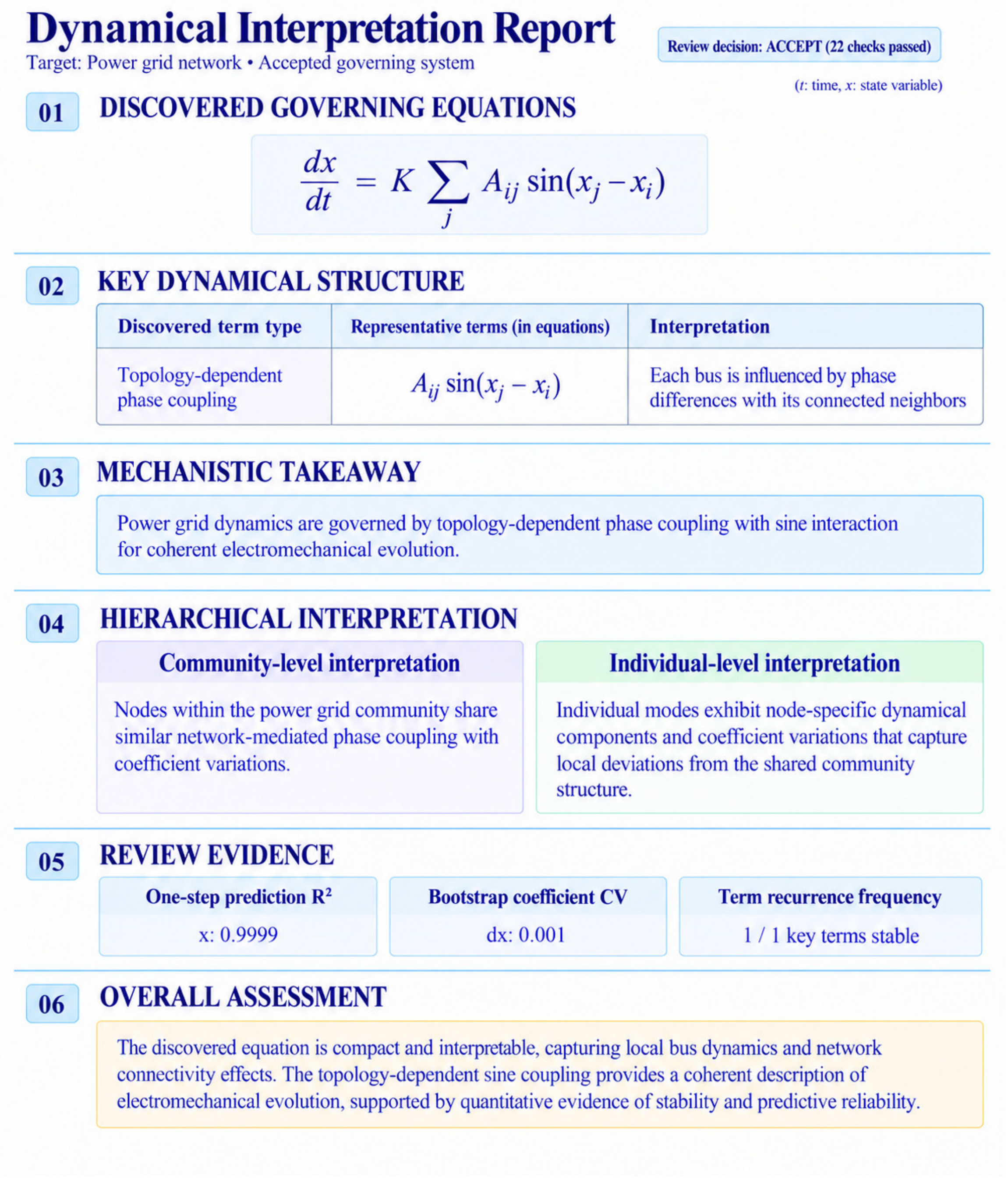

**Figure S4. Dynamical interpretation of the governing equations discovered for the power grid network.** The inferred power-grid dynamics are characterized by topology-dependent phase coupling, in which the evolution of each node is governed by the phase differences between electrically connected neighbors through a sine-coupling term. This structure captures the network-mediated synchronization mechanism underlying electromechanical dynamics. At the community level, nodes share a common phase-coupling structure, whereas individual nodes exhibit coefficient variations that reflect local dynamical heterogeneity. The retained terms show high recurrence and predictive stability, supporting the robustness of the discovered mechanism.

### *3.4 Urban rail transit network*

We use an urban rail transit passenger-flow dataset from Beijing, China. The dataset was collected from the Automatic Fare Collection (AFC) system between 05:00 and 23:00 over five consecutive weeks, from February 29 to April 3, 2016. The network contains 17 lines and 276 transit stations. Each station is treated as a dynamical node, and its passenger-flow state is represented by two variables. For station $i$, the state vector is defined as $\boldsymbol{x}_i(t) = [x_i(t), y_i(t)]^{\mathrm{T}}$, where $x_i(t)$ and $y_i(t)$ represent the entry and exit passenger flows, respectively. The passenger-flow data are aggregated at a 15-minute temporal resolution, providing sequential observations for governing equation discovery.

For network construction, the adjacency matrix $\boldsymbol{A}_{ij}$ is constructed based on the physical connectivity of the transit network, where two stations $i$ and $j$ are considered connected if they are directly linked by a rail segment. Accordingly, $\boldsymbol{A}_{ij} = 1$ indicates a direct physical connection between stations $i$ and $j$, whereas $\boldsymbol{A}_{ij} = 0$ otherwise. The resulting topology characterizes the physical connectivity through which passenger flow dynamics may interact across neighboring stations and is jointly used with the station-level state trajectories for hierarchical governing-equation discovery. Detailed governing-equation inference and dynamical interpretation results are presented below.

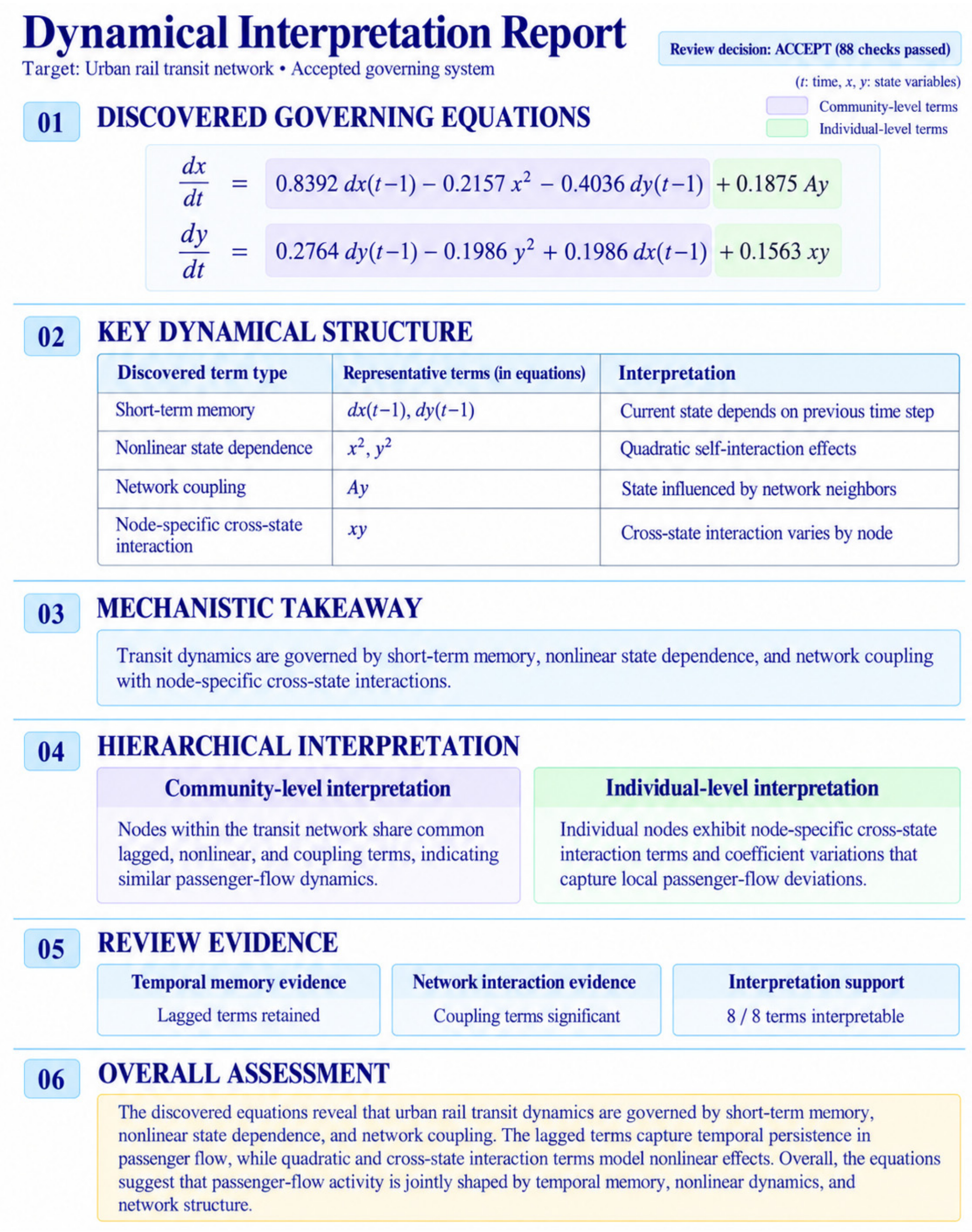

**Figure S5. Dynamical interpretation of the governing equations discovered for the urban rail transit network.** The discovered passenger-flow dynamics combine short-term temporal memory, nonlinear state dependence, network coupling, and node-specific cross-state interactions. Lagged derivative terms capture temporal persistence in passenger-flow evolution, quadratic terms describe nonlinear self-regulation, and the topology-dependent coupling term represents the influence of neighboring stations. The hierarchical decomposition shows that shared temporal, nonlinear, and network mechanisms operate at the community level, while cross-state interactions and coefficient variations account for station-specific deviations. The retained terms provide consistent temporal and network interaction evidence and remain physically interpretable.

## 4. Supplementary Validation Results for Real-World Complex Networks

Figure S6 provides detailed validation results for the four real-world complex networks, further comparing the true and inferred dynamics across different state variables and their corresponding community-level and individual-level components.

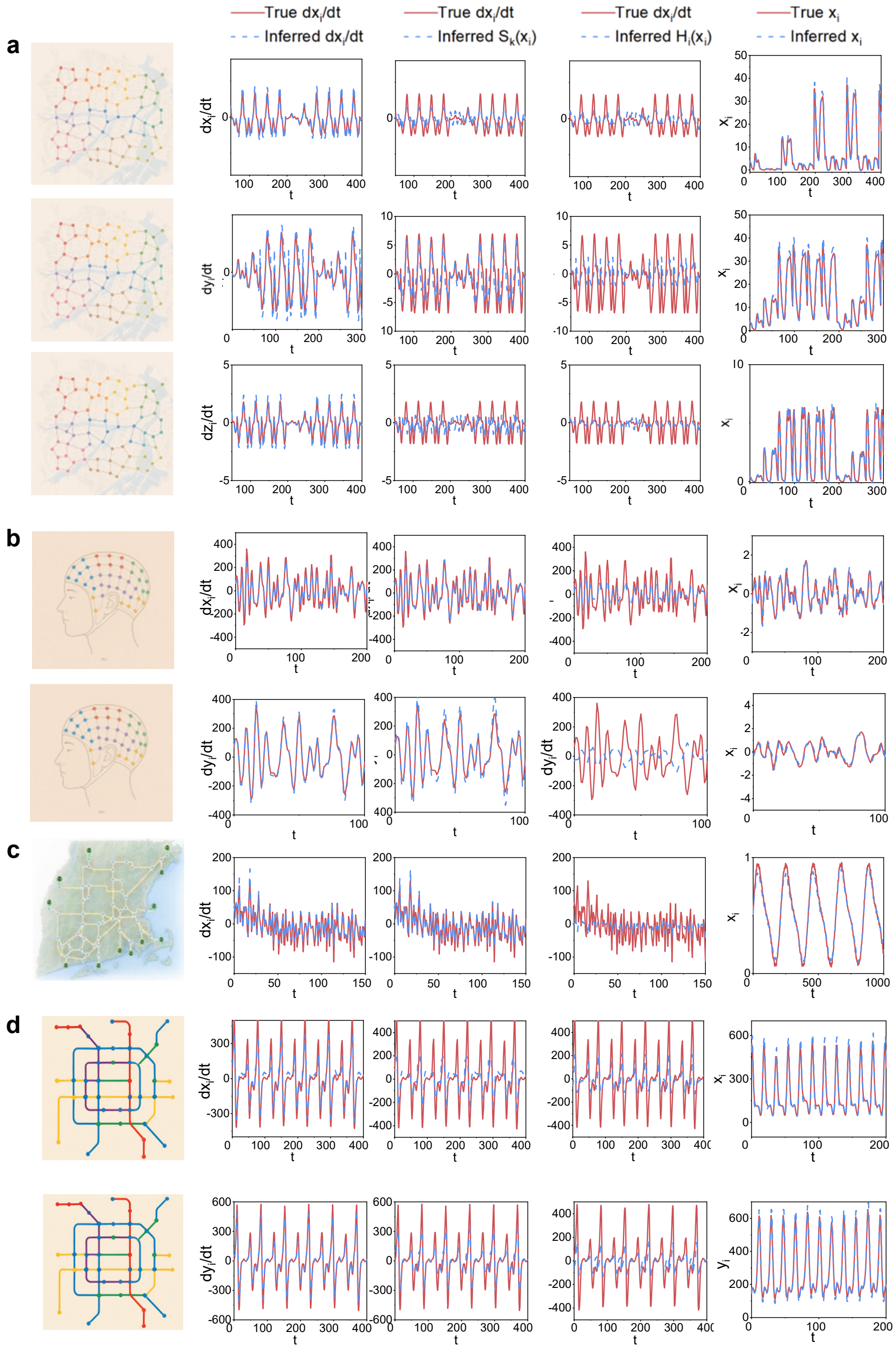

**Figure S6. Detailed validation of SAGE on four real-world complex networks.** True and inferred dynamics are compared across different state variables, together with the corresponding community-level dynamics $S_k(x)$, individual-level dynamics $H_i(x)$, and reconstructed node states. **a)** Urban mobile communication network. **b)** Brain EEG network. **c)** Power grid network. **d)** Urban rail transit network.